\documentclass{emulateapj}

\usepackage{latexsym}
\usepackage{mathrsfs}
\usepackage{amsfonts}
\usepackage{amssymb}
\usepackage{amsmath}
\usepackage{subfig}

\shorttitle{Coronal Heating}
\shortauthors{Rappazzo et al.}

\begin{document}
\title{Nonlinear Dynamics of the Parker Scenario for Coronal Heating}
\submitted{accepted by The Astrophysical Journal}

\author{A.F. Rappazzo\altaffilmark{1}, M.~Velli\altaffilmark{2}}
\affil{Jet Propulsion Laboratory, California Institute of
                        Technology, Pasadena, CA  91109, USA}

\author{G.~Einaudi\altaffilmark{3}}
\affil{Dipartimento di Fisica ``E.~Fermi'', Universit\`a di Pisa,
                        56127 Pisa, Italy} 

\and

\author{R.B.~Dahlburg\altaffilmark{4}}
\affil{Laboratory for Computational Physics and Fluid Dynamics, \\
         Naval Research Laboratory, Washington, DC 20375, USA}

\altaffiltext{1}{NASA Postdoctoral Fellow; rappazzo@jpl.nasa.gov}
\altaffiltext{2}{Also at Dipartimento di Astronomia e Scienza dello Spazio, 
                        Universit\`a di Firenze, 50125 Florence, Italy; mvelli@jpl.nasa.gov}
\altaffiltext{3}{einaudi@df.unipi.it}
\altaffiltext{4}{rdahlbur@lcp.nrl.navy.mil}
      
\begin{abstract}
The Parker or field line tangling model of coronal heating is studied comprehensively via long-time high-resolution simulations of the dynamics of a coronal loop in cartesian geometry within the framework of reduced magnetohydrodynamics (RMHD).  
Slow photospheric motions induce a Poynting flux which
saturates by driving an anisotropic turbulent cascade dominated by magnetic energy.
In physical space this corresponds to a magnetic topology where magnetic field lines are barely entangled, nevertheless current sheets (corresponding to the original tangential discontinuities hypothesized by Parker) are continuously formed and dissipated.  

Current sheets are the result of the nonlinear cascade that transfers energy from 
the scale of convective motions 
($\sim 1,000\, km$) down to the dissipative scales, where it is finally converted to heat 
and/or particle acceleration.
Current sheets constitute the dissipative structure of the system,
and the associated magnetic reconnection gives rise to impulsive ``bursty''
heating events at the small scales. This picture is consistent with the slender loops observed
by state-of-the-art (E)UV and X-ray imagers which, although apparently quiescent, shine
bright in these wavelengths with little evidence of entangled features.

The different regimes of weak and strong MHD turbulence that develop, and their influence on
coronal heating scalings, are   shown to depend on the loop parameters, and this dependence
is quantitatively characterized: weak turbulence regimes and steeper spectra occur in {\it stronger loop fields} and lead to {\it larger heating rates} than in weak field regions.

\end{abstract}

\section{INTRODUCTION}
In a previous letter \citep{rap07} we described simulations, within the framework of RMHD in cartesian geometry, aimed at solving the Parker field-line tangling (coronal heating) problem \citep{park72,park88}. We also developed a phenomenological model for nonlinear interactions, taking into account the inertial photospheric line-tying effect, which explained how the average coronal heating rate would depend on the only free parameter present in the simulations, namely the ratio of the coronal loop Alfv\'en crossing time and the photospheric eddy turnover time. This paper is devoted to a more detailed discussion of the numerical simulations and of the relationship between this work, the original Parker conjecture, and the nanoflare scenario of coronal heating.

Parker's book \citep{park94} is devoted to an examination of the basic theorem of magnetostatics,  namely that  the lowest available energy state of a magnetic field in an infinitely conducting fluid contains  surfaces of tangential discontinuity, or current sheets. It is Parker's conjecture that the continuous footpoint displacement of coronal magnetic field lines must lead to the development of such  discontinuities as the field continuously tries to relax to its equilibrium state, and it is the dynamical interplay of energy accumulation
via footpoint motion and the bursty dissipation in the forming current sheets which gives rise to the phenomenon of the high temperature solar corona, heated by the individual bursts of reconnection, or nanoflares.

What then does turbulence have to do with the nanoflare heating scenario? Parker himself strongly criticizes the use of the ``t'' word, the formation of the current sheets being due in his opinion to the  ``requirement for ultimate static balance of the Maxwell stresses''. But what  better way is there to describe the nonlinear global dynamics of a magnetically dominated 
plasma in which the formation of an equilibrium state containing current sheets 
is the inevitable asymptotic state (once the photospheric driver is turned off)? 

The striving of the global magnetic field toward a state containing current sheets must occur through local violations of the force-free condition, the induction of local flows, the collapse of the currents into ever thinner layers: a nonlinear process generating ever smaller scales. From the spectral point of view, a power law distribution of energy as a function of scale is expected, even though the kinetic energy is much smaller than the magnetic energy. The last two statements are clear indications that the word turbulence provides a correct description of the dynamical process. 

A final important issue is whether the overall dissipated power tends to a finite value as the 
resistivity and viscosity of the coronal plasma become arbitrarily small. That this must be the 
case is easy to understand (see \S~\ref{sec:eod}). For suppose that for an arbitrary, continuous, foot-point 
displacement the coronal field were only to map the foot-point motion, and that there were 
no non-linear interactions, i.e.\ the Lorentz force and convective derivatives were negligible 
everywhere. In this case, the magnetic field and the currents in the corona would then grow 
linearly in time, until the coronal dissipation at the scale of photospheric motions balanced 
the forcing. The amplitudes of the coronal fields and currents  would then be inversely 
proportional to resistivity (eqs.~(\ref{eq:bdiff})-(\ref{eq:jdiff})), and the dissipated power, 
product of resistivity and square of the current, would also scale as the inverse power of the resistivity
(eq.~(\ref{eq:hrdiff})). 
In other words, the smaller 
the resistivity in the corona, the higher the power dissipated would be. But the amplitudes 
can not become arbitrarily large, because non-linear effects intervene to stop the increase in 
field amplitudes, increasing the effective dissipation at a given resistivity. Since the power 
can not continue to increase monotonically as the resistivity is decreased, it is clear that at 
some point non-linear interactions must limit the dissipated power to a finite value, regardless 
of the value of the resistivity. Finite dissipation at arbitrarily small values of dissipative 
coefficients is another definition of a turbulent system.

All this assuming that a statistically stationary state may be reached in a finite time, a 
question closely related to the presence of finite time singularities in 3D 
magnetohydrodynamics. It now appears that magnetic field relaxation in an unforced 
situation does not lead to the development of infinitely thin current sheets in a finite time, 
but rather the current development appears to be only exponential in time \citep{GRM00}. 
In forced numerical simulations, as the ones we will describe in detail here, this is a moot 
point: for all intents and purposes a statistically stationary state is achieved at a finite time 
independent of resistivity for sufficiently high resolution. 
In fact, even if the growth is exponential, we can estimate that the width of the current sheets
reaches the meter-scale in a few tens Alfv\'en crossing times $\tau_{\mathcal{A}}$.
A typical value is $\tau_{\mathcal{A}} = 40\, s$, so that this initial time is not only finite, but
also short compared with an active region timescale. Once the steady state has been 
reached this phenomenon is no longer important. The nonlinear regime is in fact 
characterized by the presence of numerous current sheets, so that while some
of them are being dissipated others are being formed, and a {\it statistical} steady state is maintained.

It therefore seems that the Parker field-line tangling scenario of coronal heating may be described as a particular instance of magnetically dominated MHD turbulence. Numerous analytical and numerical models of this process have been presented in the past, each discussing in some detail aspects of the general problem as presented above \citep{park72,park88,HP92, vb86, ber91, StUch81, GomFF92, mik89, hen96, long94, dmi99, ein96,georg98,dmi98,ein99}. 

The numerical simulations presented here bring closure to the {\it original} question as posed in cartesian geometry by Parker, starting from a uniform axial magnetic field  straddling from one boundary plane to another,  subject to continuous independent footpoint motions in either photosphere. 
This does not imply that we have fully solved the coronal heating problem as due to footpoint dragging by the photospheric velocity field. 

A number of relevant effects have been neglected: first, the field line expansion between the photosphere and corona, which, if the photospheric flux is confined to bundles in granular and supergranular network lanes, would allow the mapping of the photospheric velocity field to the coronal volume to contain discontinuities. We are presently carrying out a dedicated set of simulations to capture this effect. Second, the projection of the 3D photospheric velocity to 2D coronal base motions parallel to the photosphere  also introduces compressibility in the forcing flow, again neglected here. Third, we have considered stationary photospheric flows. The effect of a finite eddy-turnover time in the flow was considered in \cite{ein96,georg98} in 2 dimensions, and in the ``3 dimensional'' shell model calculations of 
\cite{buc07}. These showed that time-dependence does not change things substantially provided the flow pattern does not contain degenerate symmetries, a fact confirmed by shorter simulations we defer to a future paper. Finally,  we do not address the more realistic case of a single photosphere with curved coronal loops, such as the simulations presented recently by \citet{gud05}.  While this approach has  advantages when investigating the coronal loop dynamics within its coronal neighborhood, modeling a larger part of the solar corona numerically drastically  reduces  the number of points occupied by the coronal loops.  At the moment the very low resolution attainable with this kind of simulations does not allow the development of turbulence with a well-developed inertial range. The transfer of energy from the scale of convection cells $\sim 1000\, km$ toward smaller scales is inhibited, because the smaller scales are not resolved (their linear resolution is $\sim 500\, km$).
Thus, these  simulations have not been able to shed light on the detailed coronal statistical response nor on the different regimes 
which may develop and how they depend on the coronal magnetic field crossing time and the photospheric eddy turnover time.

In \S~\ref{sec:model} we introduce the coronal loop model,
whose properties are qualitatively analyzed in \S~\ref{sec:an}.
The results of our simulations are described in \S~\ref{sec:ns},
and their turbulence properties  are analyze in more detail in \S~\ref{sec:wts}.
Finally in \S~\ref{sec:disc} we summarize and discuss our results.

\section{PHYSICAL MODEL} \label{sec:model}

A coronal loop is a closed magnetic structure threaded by a  strong axial 
field, with the footpoints rooted in the photosphere.
This makes it a strongly anisotropic system, as measured by 
the relative magnitude of the Alfv\'en velocity associated with
the axial magnetic field $v_{\mathcal A} \sim 2000\ \textrm{km}\, \textrm{s}^{-1}$ 
compared to the typical photospheric velocity $u_{ph} \sim 1\ \textrm{km}\, \textrm{s}^{-1}$. 

We study the loop dynamics in a simplified Cartesian geometry, 
neglecting field line curvature, i.e. the toroidality of  loops. Our loop is  
a ``straightened out'' box,  with an  orthogonal square cross 
section of size $\ell$ (along which the x-y 
directions lie), and an axial length $L$ (along the z direction)
embedded in an axial homogeneous uniform magnetic field 
$\mathbf{B}_0 = B_0\ \mathbf{e}_z$. 
This simplified geometry allows us to perform simulations
with both high numerical resolution and long-time duration.

In \S~\ref{sec:ge} we introduce the equations used to model the dynamics, while 
in \S~\ref{sec:bc} we give the boundary and initial conditions 
used in our numerical simulations.

\subsection{Governing Equations} \label{sec:ge}

The dynamics of a plasma embedded in a strong axial magnetic field are
well described by the equations of reduced MHD (RMHD)
\citep{kp74, stra76, mont82}. 

These equations are valid for a plasma with small ratio of
kinetic to magnetic pressures, in the limit of a large loop-aspect
ratio ($\epsilon = l/L \ll 1$, $L$ being the length of the loop and $l$ being
the minor radius of the loop) and of a small ratio of poloidal to
axial magnetic field ($b_\perp /B_0 \le \epsilon$).
In dimensionless form they can be written as:
\begin{multline}
\frac{\partial \boldsymbol{u}_{\perp}}{\partial t} 
+ \left(  \boldsymbol{u}_{\perp} \cdot \boldsymbol{\nabla}_{\perp} \right)  \boldsymbol{u}_{\perp}
= - \boldsymbol{\nabla}_{\perp} \left( p + \frac{\boldsymbol{b}_{\perp}^2}{2} \right) \\
+   \left(  \boldsymbol{b}_{\perp} \cdot \boldsymbol{\nabla}_{\perp} \right) \boldsymbol{b_{\perp}} 
+  c_{\mathcal{A}} \, \frac{\partial \boldsymbol{b}_{\perp}}{\partial z}  
+ \frac{(-1)^{n+1}}{Re_n} \boldsymbol{\nabla}^{2n}_{\perp} \boldsymbol{u}_{\perp}, \label{eq:adim1}
\end{multline}
\begin{multline}
\frac{\partial \boldsymbol{b}_{\perp}}{\partial t} =  \left(  \boldsymbol{b}_{\perp} \cdot \boldsymbol{\nabla}_{\perp} \right) \boldsymbol{u_{\perp}} -  \left(  \boldsymbol{u}_{\perp} \cdot \boldsymbol{\nabla}_{\perp} \right) \boldsymbol{b _{\perp}} \\
+  c_{\mathcal{A}} \,  \frac{\partial \boldsymbol{u}_{\perp}}{\partial z} 
+ \frac{(-1)^{n+1}}{Re_n} \boldsymbol{\nabla}^{2n}_{\perp} 
\boldsymbol{b}_{\perp}, \label{eq:adim2}
\end{multline}
\begin{equation}
\boldsymbol{\nabla}_{\perp} \cdot \boldsymbol{u}_{\perp} = 0, \qquad
\boldsymbol{\nabla}_{\perp} \cdot \boldsymbol{b}_{\perp} = 0, \qquad \qquad \qquad \quad
\label{eq:adim3}
\end{equation}
 where $\mathbf{u}_{\perp}$ and $\mathbf{b}_{\perp}$ are the 
components of the velocity and magnetic fields perpendicular to the mean field, and $p$ is the
kinetic pressure. The gradient operator likewise has only components in the $x$-$y$ plane perpendicular to the axial direction $z$, i.e.\
\begin{equation}
\mathbf{\nabla}_{\perp} = \mathbf{e}_{x}\, \frac{\partial}{\partial x} +
\mathbf{e}_{y}\, \frac{\partial}{\partial y},
\end{equation}
while the dynamics in the planes is coupled to the axial direction
through the linear terms $\propto \partial_z$.

To render the equation non dimensional magnetic fields have first been expressed in velocity units by dividing by $\sqrt{4\pi \rho_0}$ (where $\rho_0$ is a density supposed homogeneous and constant), i.e.\ considering the associated Alfv\'en velocities ($b \rightarrow b/\sqrt{4\pi \rho_0}$), and then both velocity and magnetic fields have been normalized to a typical photospheric velocity $u_{ph}$; lengths and times have been expressed in units of the
perpendicular length of the computational box $\ell$ and its
related ``eddy turnover time'' $t_{\perp} = \ell / u_{ph}$.

As a result, in equations~(\ref{eq:adim1})-(\ref{eq:adim2}), 
the linear terms $\propto \partial_z$ are multiplied by the \emph{dimensionless}
Alfv\'en velocity $c_{\mathcal{A}} = v_{\mathcal A} / u_{ph} $, i.e\ the 
ratio between the Alfv\'en velocity associated with the axial magnetic field
$v_{\mathcal A} = B_0 / \sqrt{4 \pi \rho_0}$, 
and the photospheric velocity $u_{ph}$.

The index $n$ is called \emph{dissipativity}:  the diffusive terms adopted in equations~(\ref{eq:adim1})-(\ref{eq:adim2})
correspond to ordinary diffusion for $n=1$ and to so-called hyperdiffusion
for $n > 1$. When $n=1$ the $\mathbf{\nabla}_{\perp}^2 / Re$
diffusive operator is recovered, so that $Re_1 = Re = Re_m$ corresponds 
to the kinetic and magnetic Reynolds number (considered of equal and 
uniform value):
\begin{equation} \label{r1}
Re = \frac{\rho_0\, \ell u_{ph}}{\nu}, \qquad
Re_m = \frac{4\pi \rho_0 \, \ell u_{ph} }{\eta c^2},
\end{equation}
where viscosity $\nu$ and resistivity $\eta$ are taken to be constant and
uniform ($c$ is the speed of light).

We have performed numerical simulations with both $n=1$ and $n=4$.
Hyperdiffusion is used because with a limited resolution
the diffusive timescales associated with ordinary diffusion are small enough
to affect the large scale dynamics and render very difficult the resolution 
of an inertial range, even with a grid with 512x512 points in the x-y plane
(the highest resolution grid we used for the plane).
The diffusive time $\tau_n$ at the scale $\lambda$ associated with the dissipative terms used in~(\ref{eq:adim1})-(\ref{eq:adim2}) is given by:
\begin{equation} \label{eq:taud}
\tau_n \sim Re_n\, \lambda^{2n}
\end{equation}
While for $n=1$ the diffusive time decreases relatively slowly towards smaller
scales, for $n=4$ it decreases far more rapidly. This allows to have longer
diffusive timescales at large spatial scales and similar diffusive timescales
at the resolution scale.
Numerically we require that the diffusion time at the resolution scale
$\lambda_{min} = 1 / N$, where N is the number of grid points, to be of the 
same order of magnitude for both normal and hyperdiffusion, i.e.\
\begin{equation}
\frac{Re_1}{N^2} \sim \frac{Re_n}{N^{2n}} \quad \longrightarrow \quad Re_n \sim Re_1\, N^{ 2(n-1) }
\end{equation}
For instance a numerical grid with $N=512$ points which requires a Reynolds number 
$Re_1 = 800$ with ordinary diffusion, can implement $Re_4 \sim 10^{19}$, removing
diffusive effects at the large scales, and allowing (if present) the resolution of an 
inertial range.

The numerical integration of the RMHD equations~(\ref{eq:adim1})-(\ref{eq:adim3}) 
is substantially simplified by using the potentials of the velocity ($\varphi$) and 
magnetic field ($\psi$),
\begin{equation} \label{eq:potfi}
\mathbf{u}_\perp = \mathbf{\nabla} \times 
\left( \varphi\, \mathbf{e}_z \right), \qquad
\mathbf{b}_\perp = \mathbf{\nabla} \times 
\left( \psi\, \mathbf{e}_z \right),
\end{equation}
linked to  vorticity and current by 
$\omega =- \mathbf{\nabla}^2_\perp \varphi$ and
$j = - \mathbf{\nabla}^2_\perp \psi$.

We solve numerically equations~(\ref{eq:adim1})-(\ref{eq:adim3}) 
written in terms of the potentials (see \cite{rap07}) in Fourier space,
i.e.\ we advance the Fourier components in the $x$-$y$ directions of the
scalar potentials $\varphi$ and $\psi$. 
Along the $z$ direction no Fourier transform is
performed so that we can impose non-periodic boundary conditions (specified 
in \S~\ref{sec:bc}), and a central second-order finite difference scheme is used.
In the $x$-$y$ plane a Fourier pseudospectral method is implemented.
Time is discretized with a third-order Runge-Kutta method.

We use a computational box with an aspect ratio of 10, which spans
\begin{equation}
0 \le x, y \le 1, \qquad 0 \le z \le 10.
\end{equation}

\subsection{Boundary and Initial Conditions} \label{sec:bc}

As boundary conditions at the photospheric surfaces 
($z=0,\ L$) we impose two independent velocity patterns, intended to
mimic photospheric motions, made up of large spatial 
scale projected convection cell flow patterns constant in time.  
The velocity potential at each boundary is given by:
\begin{multline} \label{eq:bc}
\varphi (x,y) = \frac{1}{\sqrt{ \sum_{mn} \alpha_{mn}^2}} \ \sum_{k,l}
            \frac{\ell\, \alpha_{kl}}{2\pi \sqrt{k^2+l^2}} \   \\ 
            \sin \left[ \frac{2 \pi}{\ell} \left( kx+ly \right) + 2\pi \, \xi_{kl} \right]
\end{multline}
We excite all the wave number values $(k,l) \in \mathbb{Z}^2$ included in the range
$3 \le \left( k^2 + l^2 \right)^{1/2} \le 4$, so that the resulting 
average injection wavenumber is $k_{c} \sim 3.4$, 
and the average injection scale $\ell_{c}$, the convection cell scale, is given by 
$\ell_c = \ell  / k_{c}$. $\alpha_{kl}$ and $\xi_{kl}$
are two sets of random numbers whose values range between $0$ and $1$, and
are independently chosen for the two boundary surfaces. The normalization adopted
in eq.~(\ref{eq:bc}) sets the value of the corresponding 
velocity rms (see eq.~(\ref{eq:potfi})) to $1/\sqrt{2}$, i.e.
\begin{equation} \label{eq:rmscon}
\int_0^{\ell} \! \! \! \int_0^{\ell} \mathrm{d}x\,  \mathrm{d}y \
\left( u_x^2 + u_y^2 \right)  = \frac{1}{2}
\end{equation}

At time $t=0$ no perturbation is imposed inside the computational box, 
i.e.\ $\mathbf{b}_{\perp} = \mathbf{u}_{\perp} = 0$, and
only the axial magnetic field $B_0$ is present: the subsequent dynamics are then the effect of the photospheric forcing~(\ref{eq:bc}) on the system, as
described in the following sections.

\section{ANALYSIS} \label{sec:an}

In order to clarify aspects of the linear and nonlinear properties of the RMHD system,  we provide an equivalent form of the equations~(\ref{eq:adim1})-(\ref{eq:adim3}).
In terms of the Els\"asser variables
$\mathbf{z}^{\pm} = \mathbf{u}_{\perp} \pm \mathbf{b}_{\perp}$, which bring out the
basic symmetry of the equations in terms of parallel and anti-parallel propagating Alfv\'en waves,
they can be written as
\begin{multline}
\frac{\partial \boldsymbol z^+}{\partial t}  =
- \left(  \boldsymbol z^- \cdot \boldsymbol{\nabla}_{\perp} \right) \boldsymbol{z}^+
+ c_\mathcal{A} \, \frac{\partial \boldsymbol z^+}{\partial z} \\
+ \frac{(-1)^{n+1}}{Re_n} \boldsymbol{\nabla}^{2n}_{\perp} \boldsymbol z^+
 - \boldsymbol{\nabla}_{\perp} P, \label{eq:els1}
\end{multline}
\begin{multline}
\frac{\partial \boldsymbol z^-}{\partial t}  =
- \left(  \boldsymbol z^+ \cdot \boldsymbol{\nabla}_{\perp} \right) \boldsymbol{z}^-
- c_\mathcal{A} \, \frac{\partial \boldsymbol z^-}{\partial z} \\
+ \frac{(-1)^{n+1}}{Re_n}   \boldsymbol{\nabla}^{2n}_{\perp} \boldsymbol z^-
 - \boldsymbol{\nabla}_{\perp} P, \label{eq:els2}
\end{multline}
\begin{equation}
\boldsymbol{\nabla}_{\perp} \cdot \boldsymbol{z}^{\pm} = 0,  \qquad \qquad \qquad \qquad
\qquad \qquad \qquad \ \ \,   \label{eq:els3}
\end{equation}
where $P = p + \mathbf{b}_{\perp}^2/2 $ is the total pressure, and is
linked to the nonlinear terms by incompressibility~(\ref{eq:els3}):
\begin{equation}  \label{eq:els4}
\mathbf{\nabla}_{\perp}^2 P =
- \sum_{i,j=1}^2 \Big( \partial_i z_j^- \Big) \Big( \partial_j z_i^+ \Big).
\end{equation}

In terms of the Els\"asser variables
$\mathbf{z}^{\pm} = \mathbf{u}_{\perp} \pm \mathbf{b}_{\perp}$, 
a velocity pattern $\mathbf{u}_{\perp}^{0,L}$ at upper or lower boundary surface 
becomes the constraint
$\mathbf{z}^{+} + \mathbf{z}^{-} = 2 \mathbf{u}_{\perp}^{0,L}$ at that boundary.
Since, in terms of characteristics (which in this case are simply $\mathbf{z}^{\pm}$
themselves), we can specify only the incoming wave
(while the outgoing wave is determined by the dynamics inside the computational
box), this velocity pattern  implies a reflecting condition at the top ($z=L$) and bottom ($z=0$) planes:
\begin{equation} \label{eq:bc0}
\mathbf{z^{-}}  = - \mathbf{z^{+}} + 2 \, \mathbf{u}^{0}_{\perp}
\quad \textrm{at} \ z=0,
\end{equation}
\begin{equation} \label{eq:bcL}
\mathbf{z^{+}} = - \mathbf{z^{-}} + 2 \, \mathbf{u}^{L}_{\perp}
\quad \textrm{at} \ z=L.
\end{equation}

The linear terms ($\propto \partial_z$) in equations~(\ref{eq:els1})-(\ref{eq:els2})
give rise to two distinct wave equations for the $\mathbf{z}^{\pm}$ fields,
which describe Alfv\'en waves propagating along the axial direction $z$. 
This wave propagation, which is present during both the linear and nonlinear regimes,
is responsible for the continuous energy influx on large perpendicular scales
(see eq.~(\ref{eq:bc}))  from the boundaries into the loop.
The nonlinear terms 
$\left(  \mathbf z^{\mp} \cdot \mathbf{\nabla}_{\perp} \right) \mathbf{z}^{\pm}$
are then responsible for the transport of this energy from the large scales toward the
small scales, where energy is finally dissipated, i.e.\ converted to heat and/or
particle acceleration.

A well-known important feature of the nonlinear terms in 
equations~(\ref{eq:els1})-(\ref{eq:els3}) is the absence of self-coupling,
i.e.\  only counterpropagating waves interact non-linearly, and if one of the two fields
$\mathbf{z}^{\pm}$ is zero, there are no nonlinear interactions at all.
This fact, i.e.\  that counter-propagating wave-packets may interact only
while they are crossing each other, lies at the 
basis of the so-called Alfv\'en effect \citep{ iro64, kra65}, 
which ultimately renders the nonlinear timescales longer and slows down the dynamics.

From this description three different timescales arise naturally: 
$\tau_{\mathcal A}$, $\tau_{ph}$ and $\tau_{nl}$.
$\tau_{\mathcal A} = L/v_{\mathcal A}$ is the crossing time
of the Alfv\'en waves along the axial direction $z$, i.e.\
the time it takes for an Alfv\'en wave to cover the loop length $L$.
$\tau_{ph} \sim 5~m$ is the characteristic time associated with photospheric
motions, while $\tau_{nl}$ is the nonlinear timescale.

For a typical coronal loop $\tau_{\mathcal A} \ll \tau_{ph}$. For instance
for a coronal loop long $L = 40,000~km$ and with an Alfv\'en velocity 
$v_{\mathcal A} = 2,000~km\, s^{-1}$ we obtain $\tau_{\mathcal A} = 20~s$,
which is small compared to $\tau_{ph} \sim 5~m = 300~s$. 
This is the reason we carried out simulations with a photospheric forcing constant in time (see eq.~(\ref{eq:bc})), i.e.\ for which formally $\tau_{ph} = \infty$. 

In the RMHD ordering the nonlinear timescale $\tau_{nl}$ is bigger than the 
Alfv\'en crossing time $\tau_{\mathcal A}$. As we shall 
see this ordering is maintained during our simulations and
we will give analytical estimates of the value of $\tau_{nl}$ as a function 
of the characteristic parameters of the system.

An important feature of equations~(\ref{eq:els1})-(\ref{eq:els3})
that we will use to generalize our results is that, apart from the 
Reynolds numbers, there is only one fundamental non-dimensional parameter:
\begin{equation} \label{eq:fsc}
f = \frac{\ell_c\, v_{\mathcal A}}{L\, u_{ph}}.
\end{equation}
Hence all the physical quantities which result from the dynamical evolution,
e.g.\ energy, Poynting flux, heating rate, timescales, etc., must depend on
this single parameter $f$.

\subsection{Energy Equation} \label{sec:ee}

From equations~(\ref{eq:adim1})-(\ref{eq:adim3}), with $n=1$,
and considering the Reynolds numbers equal,
the following energy equation can be derived:
\begin{equation} \label{eq:poy}
\frac{\partial }{\partial t} \left( \frac{1}{2}  \mathbf u_{\perp}^2
+ \frac{1}{2}  \mathbf b_{\perp}^2   \right)
= - \mathbf{\nabla} \cdot \mathbf{S}
  - \frac{1}{Re}  \left( \mathbf{ j }^2 + \mathbf{ \omega }^2 \right),
\end{equation}
where 
$\mathbf{S} = \mathbf{B} \times (\mathbf{u} \times \mathbf{B})$ 
is the Poynting vector. As expected the energy balance of the system 
described by eq.~(\ref{eq:poy}) is due to the competition between the
energy (Poynting) flux flowing into the computational box and the ohmic and viscous dissipation.
Integrating eq~(\ref{eq:poy}) over the whole box
the only relevant component of the Poynting vector is the component
along the axial direction $z$, because in the $x$-$y$ plane periodic boundary conditions
are used and their contribution to the Poynting flux is null. As 
$\mathbf{B} = c_{\mathcal{A}}\, \mathbf{e}_z + \mathbf{b}_{\perp}$ and
$\mathbf{u} = \mathbf{u}_{\perp}$, this is given by
\begin{equation} \label{eq:sz}
S_z = \mathbf{S} \cdot \mathbf{e}_z = - c_{\mathcal{A}} 
\left( \mathbf{u}_{\perp} \cdot \mathbf{b}_{\perp} \right).
\end{equation}
Considering that the velocity fields at the photospheric boundaries 
are given by $\mathbf{u}_{\perp}^0$ and $\mathbf{u}_{\perp}^L$,
for the integrated energy flux we obtain
\begin{equation} \label{eq:tsz}
S = 
  c_{\mathcal A} \int_{z=L} \! \mathrm{d} a\, \left( \mathbf{u}_{\perp}^L 
\cdot \mathbf{b}_{\perp}  \right)
- c_{\mathcal A} \int_{z=0} \! \mathrm{d} a\, \left( \mathbf{u}_{\perp}^0 
\cdot \mathbf{b}_{\perp}  \right).
\end{equation}
The injected energy flux therefore depends not only on the photospheric forcing
and the axial Alfv\'en velocity (which have fixed values), but also 
on the value of the magnetic fields at the boundaries, which is determined by the 
dynamics of the system inside the computational box: \emph{the injection of energy depends 
on the nonlinear dynamics which develops, and viceversa}.

The simplified topology investigated in this paper, i.e.\  a strong axial magnetic
field whose footpoints are dragged by 2D orthogonal motions 
applies to regions where emerging flux may be neglected. 
Consider the axial component of the velocity $u_z$
field carrying new magnetic field ($\mathbf{b}_{\perp}^{ef}$) into the corona. 
The associated Poynting flux is
\begin{equation} \label{eq:ef}
S_z^{ef} = \left( \mathbf{b}_{\perp}^{ef} \right)^2 u_z.
\end{equation}
This  flux component is negligible when  $S_z^{ef} < S_z$, i.e., since
all the components of the photospheric velocity fields are of the
same order, $u_z \sim u_{ph}$, when

\begin{equation} \label{eq:efc}
\left( b_{\perp}^{ef} \right)^2 <  B_0\, b_{\perp}^{turb}.
\end{equation}

In \S~\ref{sec:sca} we give an estimate of the value of the field $b_{\perp}^{turb}$
generated by the field-line dragging, and will be able to quantify for which value
of $b_{\perp}^{ef}$ the emerging flux can be neglected.

\subsection{Linear Stage} \label{sec:ls}

For $t < \tau_{nl}$ nonlinear terms can be neglected. Neglecting also the diffusion terms, 
which play no role on large scales, equations~(\ref{eq:adim1})-(\ref{eq:adim3})
reduce to two simple wave equations. Coupled with the boundary 
conditions~(\ref{eq:bc}) the solution for times longer
than the crossing time $\tau_{\mathcal{A}}$ reads:{\setlength\arraycolsep{-8pt}
\begin{eqnarray}
& &\mathbf{b}_{\perp} (x,y,z,t) =
\left[ \mathbf{u}^{L} (x,y) - \mathbf{u}^{0} (x,y) \right]\
\frac{t}{\tau_{\mathcal A}}, \label{eq:blin}\\
& &\mathbf{u}_{\perp} (x,y,z,t) = \mathbf{u}^L (x,y) \frac{z}{L} 
+ \mathbf{u}^0 (x,y) \left( 1 - \frac{z}{L} \right). \label{eq:vlin}
\end{eqnarray}
}
This shows that A) the loop velocity field is bounded by the imposed photospheric fields and b)
the magnetic field grows linearly in time,  uniform along the loop, while  \emph{mapping}
the photospheric velocity field in the perpendicular planes. Therefore, for a generic  
set of velocities $\mathbf{u}^L$ and $\mathbf{u}^0$,  the resulting magnetic
fields (\ref{eq:blin})-(\ref{eq:vlin}) give rise to non-vanishing 
forces in the perpendicular planes which grow quadratically in time,
becoming dynamically important after a certain interval \citep{buc07}. 

There exists a (singular) set of velocity forcing patterns, for which the generated 
coronal field has a vanishing Lorentz force. For simplicity consider
$\mathbf{u}^L = 0$: in terms of potentials it follows that
$\psi = -\varphi^0 \, t/\tau_{\mathcal A}$ and
$\varphi = \varphi^0 \, (1-z/L)$ (where $\mathbf{u}^0_\perp = \mathbf{\nabla} \times
\left( \varphi^0\, \mathbf{e}_z \right)$). 
In this case both $\mathbf{b}_{\perp}$ and $\mathbf{u}_{\perp}$ are
proportional to $\mathbf{\nabla}_{\perp} \times  ( \varphi^0\, \mathbf{e}_z )$.
The condition for the vanishing of nonlinear terms then becomes
\begin{equation}
\mathbf{\nabla} \left( \mathbf{\nabla}^2 \, \varphi^0 \right) \times
\mathbf{\nabla} \varphi^0 = 0, \quad \textrm{with} \quad \varphi^0 = \varphi^0 \left(x, y\right).
\end{equation}
This condition is then satisfied by those fields for which the laplacian is constant 
along the streamlines of the field.
As $\omega = - \mathbf{\nabla}^2 \, \varphi$ this is equivalent to the statement that
the \emph{vorticity is constant along the streamlines}. This condition is in general not 
verified, unless very symmetric functions are chosen, e.g. in cartesian geometry by any 1D 
function like $\varphi^0 = f(x)$, and in polar coordinates by any radial function $\varphi^0 = g(r)$.

Generally speaking even in such peculiar configurations non-linear interactions will arise due to
the onset of instabilities. We defer discussion of these extreme examples to a subsequent paper, the 
random photospheric fields ~(\ref{eq:bc}) discussed here always giving rise to non-vanishing forces .

Inserting the linear evolution fields  ~(\ref{eq:blin}) in the expression for the integrated energy flux~(\ref{eq:tsz}),
we find
\begin{equation} \label{eq:slin}
S = c_{\mathcal A} \int\! \mathrm{d} a\, | \mathbf{u}^L - \mathbf{u}^0 |^2 
\cdot \frac{t}{\tau_{\mathcal A}},
\end{equation}
i.e.\ the Poynting flux $S$ grows linearly in time until such a time that non-linear interactions
set in.

A similar linear analysis was already performed by \cite{park88}, who noted that if this
is the mechanism responsible for coronal heating, then the energy flux $S_z \sim S / \ell^2$ 
must approach the value $S_z \sim 10^7\, erg\, cm^2\, s^{-1}$ necessary to sustain an 
active region before a saturating mechanism, 
magnetic reconnection of singular current sheets in Parker's
language, takes over.

In fact however the value reached by $S_z$ depends on the nonlinear dynamics, 
its value \emph{self-consistently} determined by solving the \emph{nonlinear problem}. 
An $S_z$  too small compared with observational constraints would then rule 
out the Parker model.

\subsection{Effects of Diffusion} \label{sec:eod}

The linear solution~(\ref{eq:blin})-(\ref{eq:vlin}) has been obtained without
taking into account the diffusive terms. This is justified, given the large
value of the Reynolds numbers for the solar corona. But numerically it 
can be important. At very low resolution diffusion is so important
that little or no nonlinear dynamics develop and the system
reaches a balance between the photospheric forcing and
diffusion of the large scale fields.
\begin{figure}
     \includegraphics[width=0.47\textwidth]{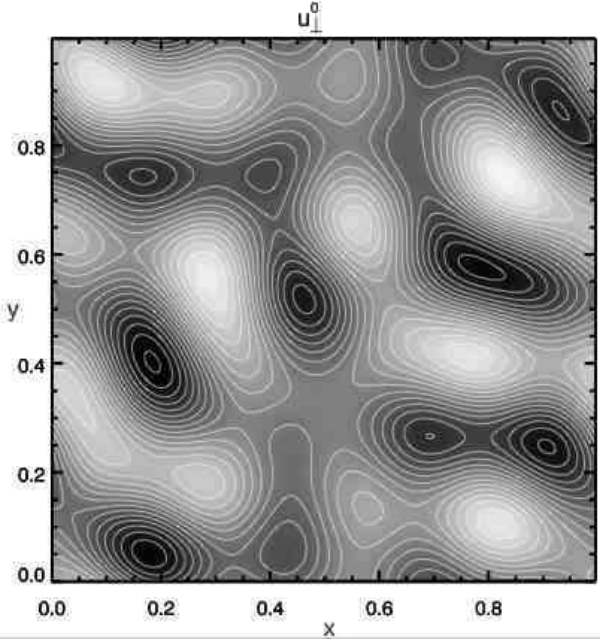}
      \caption{Streamlines of the velocity field $\mathbf{u}_{\perp}^0$,
      the boundary forcing at the \emph{bottom} plane $z=0$ for run~A. In lighter vortices the 
      velocity field is directed anti-clockwise while in darker vortices it is 
      directed clockwise. The cross-section shown in the figure is roughly
      $4000 \times 4000\, km^2$, where the typical scale of a convective
      cell is $1000\, km$.\\[0.1pt]  
     \label{fig:v0}}  
\end{figure}

One can attempt to bypass the non-linear problem by adopting a much smaller
``turbulent'' value of the Reynolds number \citep{HP92}.
For this ``ad hoc''  value of the Reynolds number  the average dissipation would 
be the same as in the high Reynolds number active turbulence limit.
Linearizing equation~(\ref{eq:adim2})  (with $n=1$ and $Re_1 = Re$), we obtain 
\begin{equation} \label{eq:lind2}
\frac{\partial \mathbf{b}_{\perp} }{\partial t} 
=  c_\mathcal{A} \frac{\partial  \mathbf{u}_{\perp}}{\partial z}
+ \frac{1}{Re} \mathbf{ \nabla}^2_{\perp}  \mathbf{b}_{\perp}.
\end{equation}
Taking into account that the forcing velocities are dominated by 
components at the injection scale $\ell_{c}$ 
(see eq.~(\ref{eq:bc})), the relation 
$\mathbf{\nabla}^2_{\perp}  \varphi = - \left( 2 \pi / \ell_{c} \right)^2 
\varphi$, where $\ell_{c} = \ell / k_{c}$ with the average wavenumber 
$k_{c} \sim 3.4$, is approximately valid.
Integrating then eq.~(\ref{eq:lind2}) over $z$ and dividing by the length $L$, 
we obtain for $\mathbf{b}_{\perp}$ averaged along $z$:
\begin{equation} \label{eq:eqdiff}
\frac{\partial \mathbf{b}_{\perp} }{\partial t} 
=  \frac{c_\mathcal{A}}{L} 
\left[ \mathbf{u}^L \left( x, y \right) - \mathbf{u}^0 \left( x, y \right) \right]
- \frac{\left( 2 \pi \right)^2}{\ell_{c}^2 Re}  \mathbf{b}_{\perp}.
\end{equation}
Indicating with $\mathbf{u}_{ph} = \mathbf{u}^L - \mathbf{u}^0$, with 
$\tau_{\mathcal{R}} = {\ell_{c}^2 Re}/ \left( 2 \pi  \right)^2$ the diffusive
time-scale and with $\tau_{\mathcal A} = L / c_{\mathcal A}$ the
Alfv\'en crossing time,  the solution is given by:
\begin{equation} \label{eq:bdiff}
\boldsymbol{b}_{\perp} \left( x, y, t \right) =
\boldsymbol{u}_{ph} \left( x, y \right)
\frac{\tau_{\mathcal R}}{\tau_{\mathcal A}}
\left[ 1 - \exp  \left( - \frac{t}{\tau_{\mathcal R}} \right) \right],
\end{equation}
\begin{multline} \label{eq:jdiff}
\left| j \left( x, y, t \right) \right| =
\left| \boldsymbol{u}_{ph} \left( x, y \right) \right|
\left( \frac{2 \pi}{\ell_{c}} \right)  
\frac{\tau_{\mathcal R}}{\tau_{\mathcal A}} \times \\
\times \left[ 1 - \exp  \left( - \frac{t}{\tau_{\mathcal R}} \right) \right].
\end{multline}
So that the magnetic energy $E_M$ 
\begin{figure}
     \includegraphics[width=0.47\textwidth]{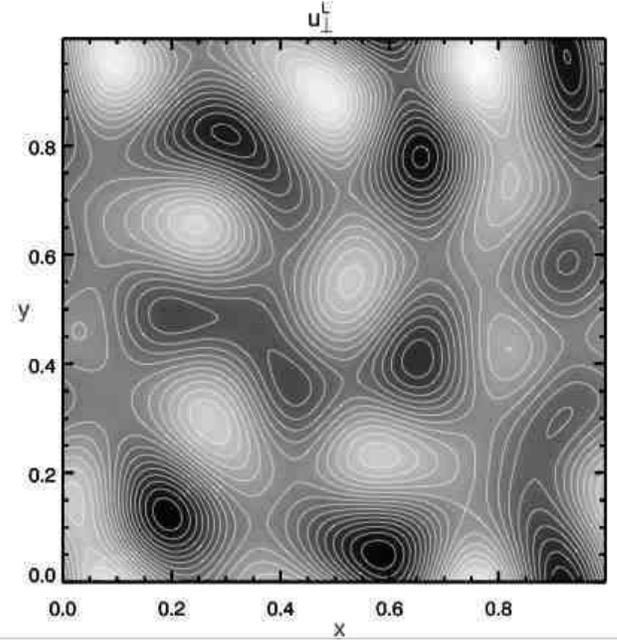}
      \caption{Streamlines of the velocity field $\mathbf{u}_{\perp}^L$,
      the boundary forcing at the \emph{top} plane $z=L$ for run~A.
      The numerical grid has 512x512 points in the x-y planes, with
      a linear resolution of $\sim 8\, km$.\\[0pt] 
     \label{fig:vL}}  
\end{figure}
and the ohmic dissipation rate $J$ are given by
\begin{multline} \label{eq:emdiff}
E_M =  \frac{1}{2} \int_V \mathrm{d}^3 \boldsymbol{x} \ \boldsymbol{b}_{\perp}^2 = \\
= \frac{1}{2} \ell^2\, L\, u^2_{ph}\,
\left( \frac{\tau_{\mathcal R}}{\tau_{\mathcal A}} \right)^2
\left[ 1 - \exp  \left( - \frac{t}{\tau_{\mathcal R}} \right) \right]^2,
\end{multline}
\begin{multline} \label{eq:hrdiff}
J =  \frac{1}{Re} \int_V \mathrm{d}^3 \boldsymbol{x} \ j^2 =  \\
= \ell^2\, L\, u^2_{ph}\,
\frac{\tau_{\mathcal R}}{\tau_{\mathcal A}^2}
\left[ 1 - \exp  \left( - \frac{t}{\tau_{\mathcal R}} \right) \right]^2,
\end{multline}
where $u_{ph}$ is the rms of $\mathbf{u}_{ph}$, and with the rms of the
boundary velocities $\mathbf{u}^0$ and $\mathbf{u}^L$ fixed to $1/2$
(\ref{eq:rmscon}) we have $u_{ph} \sim 1$.
Both total magnetic energy~(\ref{eq:emdiff}) and ohmic dissipation~(\ref{eq:hrdiff}) 
grow quadratically in time for time smaller than the resistive time $\tau_{\mathcal R}$, 
while on the diffusive time scale they saturate to the values
\begin{equation} \label{eq:satdiff}
E_M^{sat} = \frac{\ell^6\, c_{\mathcal{A}}^2\, u^2_{ph}\, Re^2}{
L\, \left( 2\pi k_{c} \right)^4}, \qquad
J^{sat} = \frac{\ell^4\, c_{\mathcal{A}}^2\, u^2_{ph}\, Re}{
L\, \left( 2\pi k_{c} \right)^2},
\end{equation}
written explicitly in terms of the loop parameters and Reynolds number.

Magnetic energy saturates to a value proportional to the square
of both the Reynolds number and the Alfv\'en velocity, while
the heating rate saturates to a value that is proportional to the Reynolds 
number and the square of the axial Alfv\'en velocity.
We have also used equations~(\ref{eq:emdiff})-(\ref{eq:hrdiff}) as a check
in our numerical simulations, and during the linear stage, before nonlinearity sets in
they are well satisfied. 

From equation~(\ref{eq:emdiff})-(\ref{eq:hrdiff}) we can estimate the saturation
time as the time at which the functions~(\ref{eq:emdiff})-(\ref{eq:hrdiff}) reach 
2/3 of the saturation values. It is approximately given by
\begin{equation} \label{eq:tdiff}
\tau^{sat} \sim 2\, \tau_{\mathcal{R}} = \frac{2\,\ell^2 Re}{\left( 2\pi k_{c} \right)^2}
\end{equation}

In the next section we describe the results of our simulations, which
investigate the linear and nonlinear dynamics. The average values may be used in conjunction with 
\ref{eq:satdiff}  to define the equivalent turbulence Reynolds number.

\section{NUMERICAL SIMULATIONS} \label{sec:ns}
\begin{table}[t]
\begin{center}
\begin{tabular}{c r l c c r}
\hline\hline
Run  &        $c_{\mathcal A}$        &  $n_x \times n_y \times n_z$
& n  &        $Re$,  $Re_4$           & $t_{max}/\tau_{\mathcal A}$ \\
\hline
A   &    200   &   $512 \times 512 \times 200$   &   1    &  $8\cdot10^{2}$    &   548 \\
B   &    200   &   $256 \times 256 \times 100$   &   1    &  $4\cdot10^{2}$    &  1061 \\
C   &    200   &   $128 \times 128 \times 100$   &   1    &  $2\cdot10^{2}$    &  2172 \\
D   &    200   &   $128 \times 128 \times 100$   &   1    &  $1\cdot10^{2}$    &   658 \\
E   &    200   &   $128 \times 128 \times 100$   &   1    &  $1\cdot10^{1}$    &  1272 \\
F   &     50   &   $512 \times 512 \times 200$   &   4    &  $3\cdot10^{20}$   &   196 \\
G   &    200   &   $512 \times 512 \times 200$   &   4    &    $10^{19}$       &   453 \\
H   &    400   &   $512 \times 512 \times 200$   &   4    &    $10^{20}$       &    77 \\
I   &   1000   &   $512 \times 512 \times 200$   &   4    &    $10^{19}$       &   502 \\
\hline
\hline
\end{tabular}
        \caption{Summary of the simulations. 
        $c_{\mathcal A}$ is the axial Alfv\'en velocity and
        $n_x \times n_y \times n_z$ is number of points for the numerical grid.
        n is the \emph{dissipativity}, $n=1$ indicates normal diffusion, $n=4$ hyperdiffusion. 
        $Re$ ($=Re_1$) or  $Re_4$ 
        indicates respectively the value of the Reynolds number or of the hyperdiffusion 
        coefficient (see eq.(\ref{eq:els1})-(\ref{eq:els2})). The duration of the simulation 
        $t_{max}/\tau_{\mathcal A}$ 
        is given in Alfv\'en crossing time unit $\tau_{\mathcal A} = L / v_{\mathcal A}$.
        \label{tab:r}}
\end{center}
\end{table}

In this section we present a series of numerical simulations, summarized in 
Table~\ref{tab:r}, modeling a coronal layer driven by a forcing velocity pattern 
\emph{constant in time}.
On the bottom and top planes we impose two independent velocity forcings
as described in \S~\ref{sec:bc}, which result from the linear combination
of large-scale eddies with random amplitudes, normalized so that
the rms of the photospheric velocity is $u_{ph} \sim 1\, km\, s^{-1}$.
For each simulation a different set of random amplitudes is chosen, corresponding
to different patterns of the forcing velocities. A realization of this forcing with 
a specific choice (run~A) of the random amplitudes is shown in Figures~\ref{fig:v0}-\ref{fig:vL}.

The length of a coronal section is taken as the unitary length.
As we excite all the wavenumbers between 3 and 4, and the
typical convection cell scale is $\sim 1,000\, km$, this implies
that each side of our  section is roughly $4,000\, km$ long.
Our typical grid for the cross-sections has 512x512 grid points,
corresponding to $\sim 128^2$ points per convective cell,
and hence a linear resolution of $\sim 8~km$. 

Between the top and bottom plates a uniform magnetic field 
$\mathbf{B} = B_0\, \mathbf{e}_z$ is present. The subsequent
evolution is due to the shuffling of the footpoints of the magnetic
field lines by the photospheric forcing.

In the different numerical simulations, keeping fixed the cross-section
length ($\sim 4,000\, km$) and axial length ($\sim 40,000\, km$),
we explore the behavior of the system for different values of 
$c_{\mathcal A}$, i.e.\ the ratio between the Alfv\'en velocity associated with
the axial magnetic field and the rms of the photospheric motions (density is 
supposed uniform and constant). 

Nevertheless, as shown in (\ref{eq:fsc}) the fundamental parameter is
$f = \ell_c v_{\mathcal{A}} / L u_{ph}$, so that changing 
$c_{\mathcal A} = v_{\mathcal{A}} / u_{ph}$
is equivalent to explore the behaviour of the system for different
values of $f$, where the same value of $f$ can be realized with a different
choice of the quantities, provided that the RMHD approximation is valid,
i.e.\ we are describing a slender loop threaded by a strong magnetic field.

We also perform simulations with different numerical resolutions, i.e.\ different Reynolds
numbers, and both normal ($n=1$) and hyper-diffusion ($n=4$).

The qualitative behaviour of the system is the same for all the simulations
performed. In the next section we describe these qualitative features in
detail for run~A, and then describe the quantitative differences found in 
the other simulations.

\subsection{Run~A} \label{sec:runA}

In this section we present the results of a simulation performed with a numerical
grid with 512x512x200 points, normal ($n=1$) diffusion with a Reynolds number
$Re=800$, and the Alfv\'en velocity $v_{\mathcal A}  = 200\, km\, s^{-1}$
corresponding to a ratio $c_{\mathcal A} = v_{\mathcal A}/u_{ph} = 200$. 
The streamlines of the forcing velocities
applied in the top ($z=L$) and bottom ($z=0$) planes are shown in 
Figures~\ref{fig:v0}-\ref{fig:vL}.
The total duration is roughly 550 axial Alfv\'en crossing times
($\tau_{\mathcal A} = L / v_{\mathcal A}$).

\begin{figure}
     \includegraphics[width=0.47\textwidth]{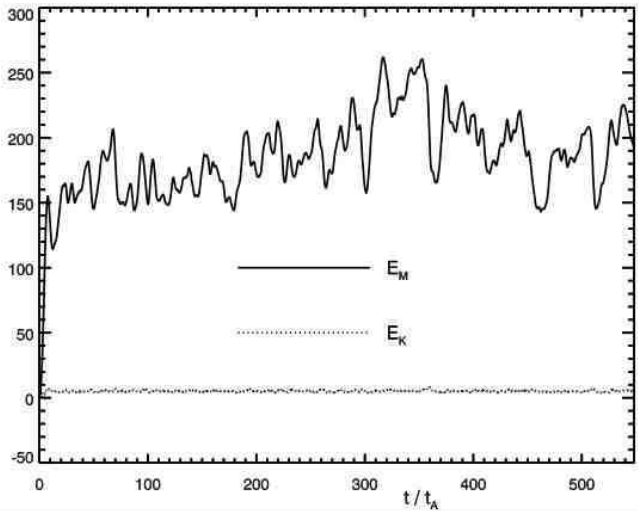}
      \caption{\emph{Run A}: High-resolution simulation with
               $v_{\mathcal A}/u_{ph} = 200$, 512x512x200
               grid points and $Re=800$.
               Magnetic ($E_M$) and kinetic ($E_K$) energies as a function
               of time ($\tau_{\mathcal A}=L / v_{\mathcal A}$ is the axial 
               Alfv\'en crossing time).\\[0pt] 
     \label{fig:fig3}}  
\end{figure}

\begin{figure}
     \includegraphics[width=0.47\textwidth]{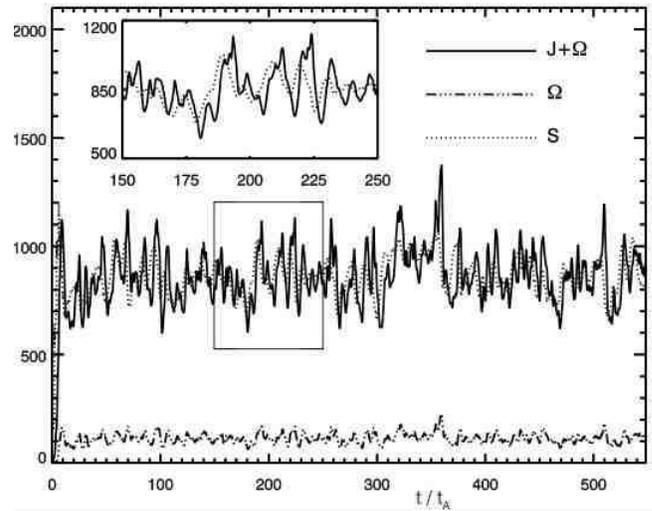}
      \caption{\emph{Run A}: The integrated Poynting flux $S$ dynamically 
               balances the Ohmic ($J$) and viscous ($\Omega$) dissipation.
               Inset shows a magnification of total dissipation and
               $S$ for $150 \le t/\tau_{\mathcal{A}} \le 250$.\\[0pt]
     \label{fig:diss}}  
\end{figure}

Plots of the total magnetic and kinetic energies
\begin{equation} \label{eq:em}
E_M = \frac{1}{2} \int\! \mathrm{d} V\, \mathbf{b}_{\perp}^2,
\qquad
E_K = \frac{1}{2} \int\! \mathrm{d} V\, \mathbf{u}_{\perp}^2,
\end{equation}
and of the total ohmic and  viscous dissipation rates
\begin{equation} \label{eq:tdiss}
J = \frac{1}{Re} \int\! \mathrm{d} V\, \mathbf{j}^2,
\qquad
\Omega = \frac{1}{Re} \int\! \mathrm{d} V\, \mathbf{\omega}^2,
\end{equation}
along with the incoming energy rate (integrated Poynting flux) $S$ (see eq.~(\ref{eq:tsz})),
are shown in Figures~\ref{fig:fig3}-\ref{fig:diss}. At the beginning the 
system has a linear behavior (see eqs.~(\ref{eq:blin})-(\ref{eq:vlin}), and 
(\ref{eq:slin})), characterized by a linear growth in time for the magnetic energy, 
the Poynting flux and the electric current, which implies a quadratic 
growth for the ohmic dissipation  $\propto (t/\tau_{\mathcal A})^2$, until time
$t \sim 6\, \tau_{\mathcal A}$, when nonlinearity sets in.
We can identify this time as the nonlinear timescale, i.e.\ 
$\tau_{nl} \sim 6\, \tau_{\mathcal A}$. The timescales of the system will be 
analyzed in more details in \S\ref{sec:tmsc}.

After this time, in the fully nonlinear stage, a \emph{statistically steady state} 
is reached, in which the Poynting flux, i.e.\ the energy that is entering
the system for unitary time, balances on time average the total dissipation
rate ($J+\Omega$). As a result there is no average accumulation of energy 
in the box, beyond what has been accumulated during the linear stage,
while a detailed examination of the dissipation time series (see inset in 
Figure~\ref{fig:diss}) shows that the Poynting flux and total dissipations
are decorrelated around dissipation peaks.

In the diffusive case from eqs.~(\ref{eq:emdiff})-(\ref{eq:tdiff}), with
the values of this simulation we would obtain $\tau^{sat} \sim 50\, \tau_{\mathcal A}$,
$E_M^{sat} \sim 6100$ and $J^{sat} \sim 7100$; all values well beyond 
those of the simulation. A value of $Re=85$ would fit the simulated
average dissipation, while $Re=140$ would approximately fit the average magnetic energy.
In any case this would only fit the curves, but \emph{the physical phenomena 
would be completely different}, as we describe in the following sections.

\emph{An important characteristic of the system is the magnetic predominance 
for both energy and dissipation} (Figures~\ref{fig:fig3} and~\ref{fig:diss}). 
In the linear stage (\S~\ref{sec:ls}) while the magnetic
field grows linearly in time, the velocity field does not, and its
value is roughly the sum of the boundary forcing fields. 
The physical interpretation is that because we are bending
the axial magnetic field with a constant forcing, as a result the 
perpendicular magnetic
field grows linearly in time, while the velocity remains limited.
More formally this is a consequence of the fact that, while on the perpendicular
magnetic field no boundary condition is imposed, the velocity field
must approach the imposed boundary values at the photosphere both
during the linear and nonlinear stages.

\begin{figure}
     \includegraphics[width=0.47\textwidth]{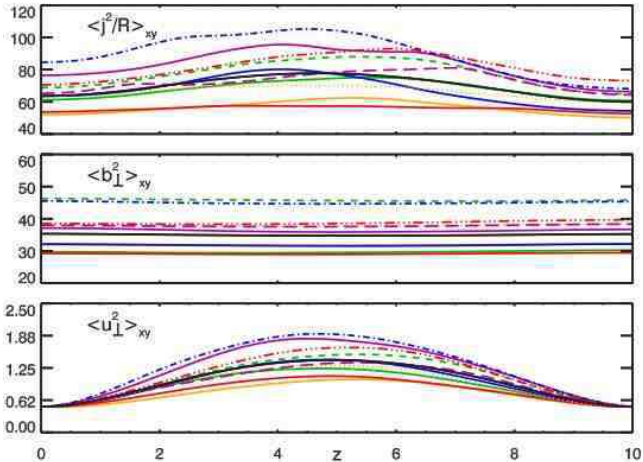}
      \caption{\emph{Run A}: 2D averages in the x-y planes of the ohmic dissipation
                $j^2/Re$, the magnetic and velocity fields 
                $\mathbf{b}_{\perp}^2$, and $\mathbf{u}_{\perp}^2$,
                as a function of z. The different colours represent 10
                different times separated by $\Delta t = 50\, \tau_{\mathcal A}$
                in the interval 
                $30\, \tau_{\mathcal A} \le  t \le 480\, \tau_{\mathcal A}$.\\[0pt]
     \label{fig:avz}}  
\end{figure}

In Figure~\ref{fig:avz} the 2D averages in the x-y planes of the 
magnetic and velocity fields and of the ohmic dissipation $j^2/Re$,
are plotted as a function of z at different times.
These macroscopic quantities are smooth and present almost no
structure along the axial direction. The reason is that
every disturbance or gradient along the axial 
direction, at least considering the large perpendicular scales 
(for the small scales behavior see \S~\ref{sec:wts} ),
is smoothed out by the fast propagation of Alfv\'en waves along the
axial direction, their propagation time $\tau_{\mathcal A}$ is in fact
the fastest timescale present (in particular $\tau_{\mathcal A} < \tau_{nl}$),
and then the system tends to be homogeneous along
this direction. 
  
The predominance of the ohmic over the viscous dissipation
is due to the fact that, as we show in the next sections, the dissipative structures
are current sheets, where magnetic reconnection takes place.

The phenomenology described in this section is general and we have
found it in all the simulations that we have performed, in particular we have 
always found that in the nonlinear stage a statistical steady state is 
reached where energies fluctuate around a mean value and
total dissipation and Poynting flux on the average balance while
on small timescales decorrelate. In particular, to check the
temporal stability of these features, which are fully confirmed, 
we have performed a numerical simulation (run~C) with the same parameters as run~A,
but with a lower resolution (128x128x100), a Reynolds number $Re = 200$ 
and a longer duration ($t \sim 2,000\, \tau_{\mathcal A}$). On the opposite the 
average levels of the energies and of total dissipation depend 
on the parameters used as we describe in the next sections.  

Before describing these features, in the next section we describe 
the current sheets formation, their temporal evolution and other properties.

\subsubsection{Current Sheets, Magnetic Reconnection,
Global Magnetic Field Topology and Self-Organization} \label{sec:csf}

\begin{figure}
     \includegraphics[width=0.47\textwidth]{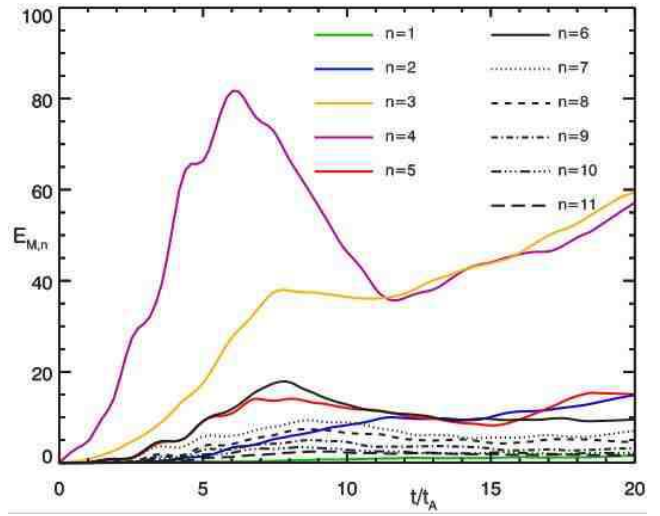}
      \caption{\emph{Run A}: First 11 magnetic energy modes as a function of time
               covering the first 20 Alfv\'en crossing times $\tau_{\mathcal{A}}$.
               Photospheric motions inject energy at $n=3$ and $4$.\\[0pt]
     \label{fig:emmod}}  
\end{figure}

The nonlinear stage is characterized by the presence of current sheets
elongated along the axial direction (Figures~(18a)-(18b)), 
which exhibit temporal dynamics and are the dissipative structures of the system.
We now show that they are the result of a nonlinear cascade. Figure~\ref{fig:emmod} 
shows the time evolution of the first 11 modes of magnetic energy for the first 
20 crossing times $\tau_{\mathcal A}$ for run~A.
During the linear stage the magnetic field is given by eq.~(\ref{eq:blin})
and is the mapping of the difference between the top ($z=10$) and bottom ($z=0$)
photospheric velocities  $\mathbf{u}^{L} (x,y) - \mathbf{u}^{0} (x,y)$,
whose streamlines are shown in Figure~17a.
The field lines of the orthogonal magnetic field in the midplane ($z=5$)
at time $t=0.63\, \tau_{\mathcal A}$ are shown in Figure~17b,
and as expected they map the velocity field. The same figure
shows in colour the axial current $j$. As shown by eq.~(\ref{eq:blin})
(taking the curl) the large scale motions that we have imposed
at the photosphere induce large scale currents in all the volume and,
as described in the previous section, if there was not a nonlinear
dynamics a balance between diffusion and forcing would be reached,
where no small scale would be formed and the magnetic field would
always map the photospheric velocities.

As time proceeds the magnetic field grows  and a cascade
transfers energy from the large scales, where the photospheric forcing (\ref{eq:bc})
injects energy at the wavenumbers $n=3$ and $4$, to the small scales (Figure~\ref{fig:emmod}).
In physical space this cascade corresponds to the collapse of the large scale currents
which lead to the formation of current sheets,
as shown in Figures~17c and 17d.
In Figures~17e and 17f we show the magnetic field lines
at time $t = 18.47\, \tau_{\mathcal A}$, in the fully nonlinear stage, with respectively
the axial component of the current $j$ and of the vorticity $\omega$.
The resulting magnetic topology is quiet complex, X and Y-points are not in fact 
easily distinguished. They are distorted and very often a component of 
the magnetic field orthogonal to the current sheet length is present, so that 
the sites of reconnection are more easily identified by the corresponding 
vorticity quadrupoles.
As shown in Figures~17e 
and 17f, the more or less distorted current sheets are always 
embedded in quadrupolar structures for the vorticity, a characteristic maintained 
throughout the whole simulation, and a  clear indication that \emph{nonlinear magnetic 
reconnection} is taking place.

Figures~18a and 18b show a view from the side
and the top of the 3D current sheets at time $t = 18.47\, \tau_{\mathcal A}$. 
When looked from the side the current sheets, 
which are elongated along the axial direction, look space filling, 
but the view from the top shows that the filling factor is actually small
(see also Figure~17).

Another aspect of the dynamics is \emph{self-organization}: while
until time $t = 4.79\, \tau_{\mathcal A}$ the magnetic field lines
are still approximately a  mapping of the photospheric velocities,
in the fully nonlinear stage they \emph{depart} from it and have an
independent  topology that evolves dynamically in time
(see the associated movie for the time evolution covering
$40$ crossing times from $\sim 508\, \tau_{\mathcal{A}}$ 
up to $\sim 548\, \tau_{\mathcal{A}}$; notations and simulation
are the same used in Figure~17).
The reason for which the photospheric forcing does not determine
the spatial shape of the magnetic field lines is due to the bigger 
value of  the rms of the magnetic field $b_{\perp} = < \mathbf{b}_{\perp}^2>^{1/2}$
in the volume respect to the rms of the photospheric forcings
$u_{ph} = < (\mathbf{u}_{\perp}^0  - \mathbf{u}_{\perp}^L)^2   >^{1/2} \sim 1$
(eqs.~(\ref{eq:bc0})-(\ref{eq:bcL})).

This means that the contribution to the dynamics of the Alfv\'enic perturbations 
propagating from the boundary are much smaller, over short periods of time, 
than the self-consistent non-linear evolution due to the magnetic fields inside 
the domain, and therefore can not determine the topology of the magnetic field.
For run~A and G, both with $c_{\mathcal{A}} = 200$, the ratio is 
$b_{\perp} / u_{ph} \sim 6$ 
and it increases up to $b_{\perp}/u_{ph} \sim 27$ in run~I with $c_{\mathcal{A}} = 1000$.
On the other hand these waves continuously transport from the boundaries the energy that
sustain the system in a magnetically dominated statistically steady state.

All the facts presented in this section, and the properties of the cascade and of 
the resulting current sheets  in presence of a magnetic guide field outlined in 
\S~\ref{sec:wts}, lead to the conclusion that the current sheets 
do not generally result directly from a ÒgeometricalÓ misalignment of neighboring magnetic 
field lines stirred by their footpoints motions, but that \emph{they are the result of a nonlinear
cascade in a self-organized system}.

Although the magnetic energy dominates over the kinetic energy,
the ratio of the rms of the orthogonal magnetic field over the
axial dominant field $B_0$ is quite small. For $c_{\mathcal{A}} = 200, 400$
and $1000$ it is $\sim 3\%$, so that the average inclination of the magnetic fieldlines
respect to the axial direction is just $\sim 2^{\circ}$, it is only for the 
lower value $c_{\mathcal{A}} = 50$ that $b_{\perp}/B_0 \sim 4\%$ and
the angle is $\sim 4^{\circ}$.
The field lines of the total magnetic
field at time $t=18.47\tau_{\mathcal A}$ are shown in 
Figures~18c and 18d. The computational box
has been rescaled for an improved viewing, and to attain the original aspect
ratio the box should be stretched 10 times along the axial direction.
The magnetic topology for the total field is quiete simple, as the line appear slightly
bended. It is only in correspondence of the small scale current-sheets that
field lines on the opposite side may show a relative inclination.
But as the current sheets are very tiny (and their width decreases at 
higher Reynolds numbers), they occupy only a very small fraction of the volume, so
that the bulk of the magnetic field lines appears only slightly bended.

It is often suggested, or implicitly assumed, that current sheets
are formed  because the magnetic field line footpoints are
subject to a \emph{random walk}. The complexity of the footpoint
trajectory would then be a necessary ingredient. In fact it would give
rise to a complex topology for the  coronal magnetic field, 
leading either to tangled field lines which would then
release energy via fast magnetic reconnection, or to turbulence.
So that the ``complexity'' of the footpoint motions would be responsible
for the ``complex'' dynamics in the corona.

On the opposite our simulations show that 
\emph{this system in inherently turbulent}, and that 
``simple'' footpoint motions give rise to turbulent dynamics
characterize by the presence of an inertial range (\S~\ref{sec:wts})
and dynamical current sheets.
In fact our photospheric forcing velocities 
(Figures~\ref{fig:v0}-\ref{fig:vL}) are constant in time and have
only large-scale components (eq.~(\ref{eq:bc})),
so that the \emph{footpoint motions} are ``ordered'' and 
\emph{do not follow any random walk}.
During the linear stage this gives rise to a magnetic field
that grows linearly in time (eq.~(\ref{eq:blin})) and that is a
mapping of the velocity fields (see eq.~(\ref{eq:blin}) and 
Figures~17a and 17b), i.e.\ both the 
magnetic field and the current have only large-scale components.
The footpoint motions of our photospheric velocities never
bring two magnetic field lines close to one another, i.e.\
they never geometrically produce a current sheet.
Current sheets are produced on an ideal timescale, the nonlinear timescale,
by the cascade.
Furthermore, as we show in the next section, the statistically steady state
that characterizes the nonlinear stage results from the \emph{balance
at the large-scales between the injection of energy and the flow of this
energy from the large scales toward the small scales}, where it is finally
dissipated.

As the system is self-organized and the magnetic energy increases
at higher values of the axial magnetic fields, very likely 
different static or time-dependent (with the characteristic photospheric time $\sim 300\, s$)
forcing functions, will not be able to determine the spatial shape of the 
orthogonal magnetic field.
In our more realistic simulation with $c_{A} =1000$ the ratio $b_{\perp} / u_{ph}$ is
in fact $\sim 27$.
Other forcing functions are currently being investigated, and time-dependent
forcing functions are likely to modulate with their associated timescale
the rms of the system, like total energy and dissipation.

\section{TURBULENCE} \label{sec:wts}

Before analyzing in detail further aspects of our simulations, namely 
inertial spectra, anisotropies and scaling laws, 
let us briefly justify the statement that the time-dependent
Parker problem, i.e.\ the dynamics of a magnetofluid threaded
by a strong axial field whose footpoints are stirred by a velocity 
field, is an MHD turbulence problem.

The fact that at the large orthogonal scales the Alfv\'en
crossing time $\tau_{\mathcal{A}}$ is the fastest timescale
so that during the linear stage the 
fields evolves as ~(\ref{eq:blin})-(\ref{eq:vlin}), means that
the photosphere's role is to contribute an anisotropic magnetic forcing
function that stirs the fluid, with an orthogonal length typical 
of the convective cells ($\sim 1000\, km$) and an
axial length is given by the loop length $L$.

Typically,  forced MHD turbulence simulations (e.g.\ see
\citet{bisk03} and references therein) are performed
using a 3-periodic numerical cube with a volumetric 
forcing function which mimics some 
physical process injecting energy at the large scales.

Solutions~(\ref{eq:blin})-(\ref{eq:vlin}) can be approximately
obtained introducing the\emph{magnetic} forcing function 
$\mathbf{F}_m$ in equation~(\ref{eq:adim2})
\begin{equation} \label{eq:mforc}
\mathbf{F}_m = 
\frac{ \mathbf{u}^{L} (x,y) - \mathbf{u}^{0} (x,y) }{
\tau_{\mathcal A}},
\end{equation}
and implementing 3-periodic boundary conditions in our elongated
($0 \le x, y \le 1$, $0 \le z \le L$) computational box.
During the linear stage this forcing would give rise, apart from the small
velocity field~(\ref{eq:vlin}), to the same magnetic field.
During the nonlinear stage, as $\tau_A < \tau_{nl}$, it
would still give rise to a similar injection of energy.
This property was the basis for the body of previous 2D calculations  \citep{ein96,dmi98,georg98}

In particular the photospheric motions imposed at the boundaries
for the Parker problem take the place of, and represent a different physical  realizations of the forcing function generally used for the 3-periodic MHD
turbulence box. In the Parker model, the equivalent forcing stirs the magnetic field, whiile in standard simulations the forcing stirs both velocity and magnetic fields or mostly the velocity field.
The main differences between ``standard'' MHD turbulence simulations
and the problem at hand are that a) the peculiarity of the low-frequency 
photospheric forcing leads to magnetic energy largely dominating over 
the kinetic energy in the system b) the forcing involves line-tying of the magnetic field
with 3-periodic boundary conditions. Line-tying inhibit the inverse cascade for the magnetic field,
as described later in this section (\S~\ref{sec:lt}). Equivalently, one may say that line-tying
hinders magnetic reconnection by rendering it less energetically favorable due to 
the increased field line-curvature it requires compared to the unbound system. This
property is fundamental to the anomalous scaling laws and enhanced
overall heating rates that will be found below.

In MHD, the cascade takes place preferentially in planes orthogonal
to the local mean magnetic field (\cite{sheb83}). The small scales 
formed are not uniformly distributed in this plane, rather they are organized in 
dynamical current-vortex sheets extended along the direction of the local 
main field. These current sheets with associated quadrupolar vorticity 
filaments \emph{form the dissipative structures of MHD
turbulence} (e.g.\ \cite{bisk00}, \cite{bisk03} and references therein). 
In our case, because the axial field is strong, the current sheets are elongated along
the axial direction to the point of being quasi-uniform along the loop axis (Figure~18).

\subsection{Spectral Properties} \label{sec:sp}

In order to investigate inertial range spectra, 
we have carried out four simulations (runs F, G, H and I in 
Table~\ref{tab:r}) with a resolution of 512x512x200 grid points
using a mild power ($n=4$) for hyperdiffusion~(\ref{eq:els1})-(\ref{eq:els2}). 

\begin{figure}
     \includegraphics[width=0.47\textwidth]{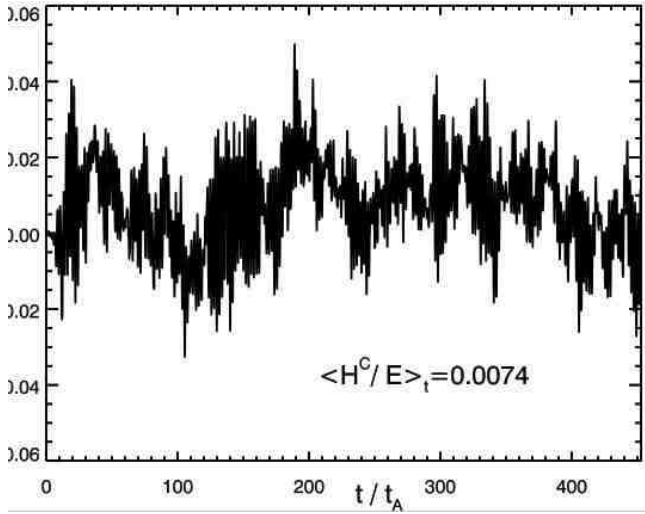}
      \caption{\emph{Run G}: Ratio between cross-helicity $H^C$ and
      total energy $E$ as a function of time. $H^C \ll E$ shows that 
      the system is in  a regime of balanced turbulence.\\[0pt]
     \label{fig:hc}}  
\end{figure}

In turbulence the fundamental physical fields are the Els\"asser variables
$\mathbf{z}^{\pm} = \mathbf{u}_{\perp} \pm \mathbf{b}_{\perp}$.
Their associated energies
\begin{equation}
E^{\pm} = \frac{1}{2} \int\! \mathrm{d} V\, \left( \mathbf{z}^{\pm} \right)^2,
\end{equation}
are linked to kinetic and magnetic energies $E_K$, $E_M$ and to the cross
helicity $H^C$
\begin{equation}
H^{C} = \frac{1}{2} \int\! \mathrm{d} V\, \mathbf{u}_{\perp} \cdot  \mathbf{b}_{\perp}
\end{equation}
by
\begin{equation}
E^{\pm} =  E_K + E_M \pm H^{C}
\end{equation}
Nonlinear terms in equations~(\ref{eq:els1})-(\ref{eq:els4}) are symmetric under the exchange
$\mathbf{z}^{+} \leftrightarrow  \mathbf{z}^{-}$, so as substantially 
are also boundary conditions~(\ref{eq:bc0})-(\ref{eq:bcL}), given that the two
forcing velocities are different but have the same rms values ($=1/\sqrt{2}$).
It is then expected that $H^C \ll E$ so that none of the two energies prevails
$E^{+} \sim E^{-} \sim E$, where $E = E_K + E_M$ is total energy. In Figure~\ref{fig:hc} 
the ratio $H^C / E$ is shown as a function of time for run~G. 
Cross helicity has a  maximum value of 5\% of 
total energy, and its time average is $\sim 1\%$,
and similar values are found for all the simulations.
Furthermore perpendicular spectra of $E$ and $E^{\pm}$ in simulations F, G, H and I,
overlap each other, so that as expected we can also assume that
\begin{equation}
\delta z^+_{\lambda} \sim \delta z^-_{\lambda} \sim \delta z_{\lambda},
\end{equation}
where $\delta z_{\lambda}$ is the rms value of the Els\"asser fields 
$\mathbf{z}^{\pm}$ at the perpendicular scale $\lambda$. 

\begin{figure}
     \includegraphics[width=0.47\textwidth]{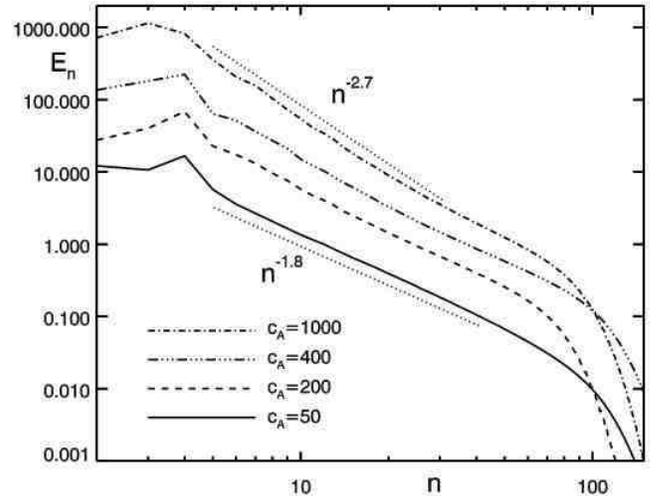}
      \caption{Total energy spectra as a function of the wavenumber $n$ for
               simulations F, G, H and I. To higher values of 
               $c_{\mathcal{A}} = v_{\mathcal{A}} / u_{ph}$,
               the ratio between the Alfv\'en and photospheric velocities, 
               correspond steeper spectra, with spectral index respectively
               $1.8$, $2$, $2.3$ and $2.7$.\\[0pt]
     \label{fig:multisp}}   
\end{figure}

In the following we always consider the spectra in the orthogonal plane $x$-$y$ 
integrated along the axial direction $z$, unless otherwise noted.
Furthermore as they are isotropic in the Fourier $k_x$-$k_y$ plane, we
will consider the integrated 1D spectra, so that for total energy
\begin{multline}
E = 
\frac{1}{2}\, \int_0^L \! \mathrm{d} z\,\iint_{0\ \ \ }^{\ell\ \, } 
\displaylimits \! \mathrm{d} x\, \mathrm{d} y\,
\left( \boldsymbol{u}^2 + \boldsymbol{b}^2  \right) = \\
= \frac{1}{2}\, \int_0^L \! \mathrm{d} z\, \ell^2 \sum_{\boldsymbol{k}}
\left( | \boldsymbol{\hat{u}} |^2  + | \boldsymbol{\hat{b}} |^2
\right) \left( \boldsymbol{k}, z \right)
= \sum_n E_n, \\
n = 1, 2, \dots  \label{eq:esp}
\end{multline}
where, similarly to eq.~(\ref{eq:bc}), $n$ indicates ``rings'' in $k$-space.
Figure~\ref{fig:multisp} shows the total energy spectra $E_n$ 
averaged in time, obtained from the hyperdiffusive 
simulations~F, G, H and I with dissipativity
$n=4$ (eqs.~(\ref{eq:els1})-(\ref{eq:els2})) and respectively 
$c_{\mathcal{A}}=50$, $200$, $400$ and $1000$.   
An inertial range displaying a power law behaviour is clearly
resolved. The spectra visibly steepens increasing the value of
$c_{\mathcal{A}}$, with spectral index ranging from $1.8$ for
$c_{\mathcal{A}}=50$ up to $\sim 2.7$ for $c_{\mathcal{A}} = 1000$.
The spectra are clearly always steeper than the well known (strong) MHD inertial range turbulence
spectra $k_{\perp}^{-5/3}$ or $k_{\perp}^{-3/2}$ . 

This steepening is certainly not a numerical artifact: 
the use of hyperdiffusion gives rise to a hump at high 
wave-number values, known
as the bottleneck effect \citep{falk94}, which when present
 \emph{flattens} the spectra. Furthermore we use the same value of dissipativity ($n=4$)
used by \cite{mg01}, who find the same IK spectral slope ($-3/2$), also confirmed in recent 
higher-resolution simulations performed by \cite{mg05} with standard $n=1$
diffusion. 

In our simulations,  a hump or flattening at high
wavenumbers is best visible in run~H with $c_{\mathcal{A}} = 400$, which might be
due to the bottleneck effect, but a more probable interpretation involves a 
transition from weak to strong turbulence at the smaller scales 
within the inertial range, which requires a preliminary 
discussion of strong vs. weal turbulence
in MHD.

\begin{figure}
     \includegraphics[width=0.47\textwidth]{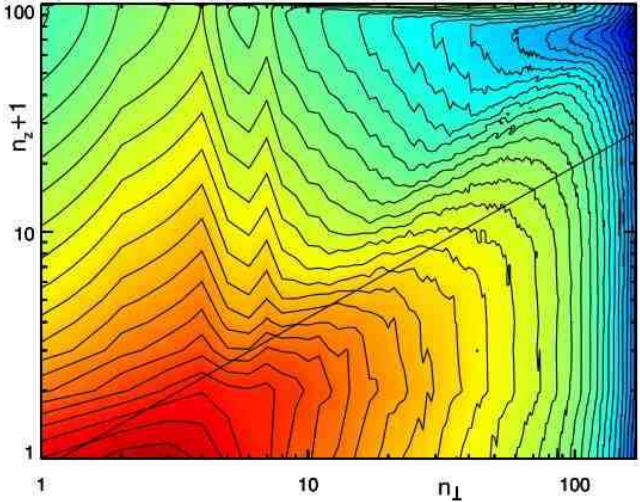}
      \caption{\emph{Run I}: Snapshot of the 2D spectrum $E(n_{\perp}, n_z)$ in 
       bilogarithmic scale at time $t \sim 145\, \tau_{\mathcal{A}}$.
       $n_{\perp}$ and $n_z$ are respectively the orthogonal and axial wavenumbers.
       The 2D spectrum is shown as a function of $n_{\perp}$ and $n_z+1$, to allow
       the display of the $n_z=0$ component.\\[0pt]
     \label{fig:spzp}}  
\end{figure}

Recently a lot of progress has been made in the understanding
MHD turbulence both in the condition of so-called strong
\citep{gs95,gs97,cv00,bisk00,mbg03,mg05,bd05,bd06,mcb06}
and weak turbulence \citep{ng97,gs97,gal00,gc06}. 
Weak turbulence has been investigated mainly through analytical
methods. The total energy spectrum can be characterized
by a $k_{\perp}^{-2}$ power law, which is easily found phenomenologically
by considering that the Alfv\'en effect occurs along the field while the cascade 
proceeds in the orthogonal direction \citep{ng97}. While our MHD simulations, 
even with our line-tying boundary conditions and anomalous
energetic regime ($b$ dominating over $u$ except at the smallest scales), 
confirm the presence of the $k_{\perp}^{-2}$ spectrum for a range of loop parameters, steeper 
spectra are also found nearly reaching $k_{\perp}^{-3}$,  
clearly linked to the strength of the axial field $B_0$ an effect we discuss more in detail in the following subsection. 

The formation of an inertial range is crucially related to the anisotropy of the cascade,
where a relationship between spectral extent in the perpendicular and parallel directions
known as ``critical balance'' may be derived. To understand this feature, consider the  
timescale $T_{\lambda}$, the energy-transfer time 
at the corresponding scale $\lambda$,
characterizing the nonlinear dynamics at that scale.
$T_{\lambda}$ does not necessarily coincide with the eddy turnover time
$\tau_{\lambda} = \lambda / \delta z_{\lambda}$ because of the Alfv\'en effect.
Spatial  structures along the axial direction result from wave propagation (at the Alfv\'en speed 
$c_{\mathcal{A}}$) of the orthogonal fluctuations. In other words,
the cascading of turbulence in two different planes separated by a distance 
 $\ell_{\parallel}$ leads to formation of  scales in the parallel direction
whose smallest size can be
\citep{gs95,cho02,ou94}
\begin{equation}
\ell_{\parallel} (\lambda) \sim c_{\mathcal{A}} T_{\lambda},
\end{equation}
the critical balance condition.
$T_{\lambda}$ will be smaller at smaller scales, so that
smaller perpendicular scales create smaller axial scales.

Figure~\ref{fig:spzp} shows a snapshot at time 
$t \sim 145\, \tau_{\mathcal{A}}$ of the 2D spectrum $E(n_{\perp}, n_z)$ for run~I 
in bilogarithmic scale, where $n_z$ and $n_{\perp}$ are respectively
the axial and orthogonal wavenumbers. Consider vertical cuts
at $n_{\perp} = const$: it is clearly visible that
from $n_{\perp} =1$ up to $n_{\perp} \sim 20$ the wavenumbers
with $n_z > 1$  are scarcely populated compared to the respective wavenumbers with $n_z \le 1$
(the parallel spectrum has also the $n_z=0$ component,
in Figure~\ref{fig:spzp} the vertical coordinate is $n_z+1$). However note also 
how the loci of maximum parallel wave-number do not precisely follow the critical 
balance line, rather they are offset at larger $n_{\perp}$: in our case,  
the hypothetical length of the axial structures (from critical balance)
can be longer than the characteristic length of the system, 
in our case the length of the coronal loop $L$. But in the
range of perpendicular wavenumbers for which 
\begin{equation}
\ell_{\parallel} (\lambda) > L,
\end{equation}
 boundary conditions, i.e. line-tying, intervenes and the cascade along the axial direction is strongly inhibited. 
In our simulations this occurs roughly at $n_{\perp} \sim 20$. Beyond $n_{\perp} \sim 20$ the spectrum is roughly constant along $n_{\perp} = const$ up to a critical value where it 
drops. 

Interestingly enough, the slope of the 1D spectrum for run~I (Figure~\ref{fig:multisp}) diminishes its value
around $n_{\perp} \sim 20$. The reason is that the condition 
$\ell_{\parallel} (\lambda) > L$ with $\ell_{\parallel} (\lambda)$ defined by critical balance,
turns out to play a major role in the ``strength'' or ``weakness'' of the cascade:
\emph{for $n_{\perp} \lesssim 20$ the system is in a weak
turbulent regime, while for $n_{\perp} \gtrsim 20$ a transition to 
strong turbulence is observed}.
\begin{figure}
     \includegraphics[width=0.47\textwidth]{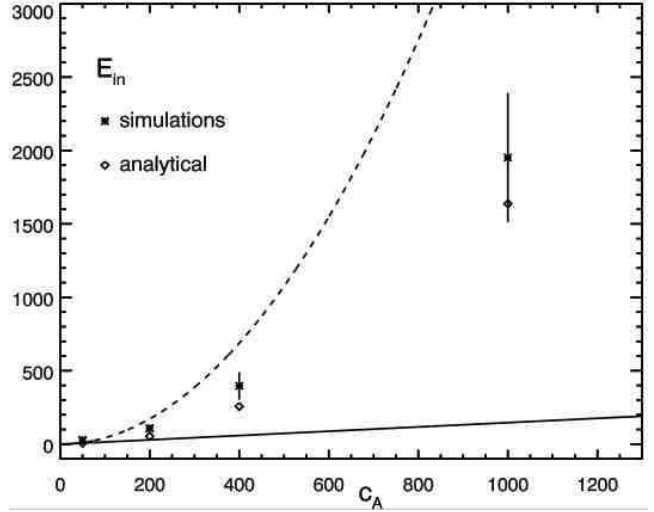}
      \caption{Total energy at the injection scale (modes 3 and 4), time-averaged
      for the four simulations F, G, H and I with different Alfv\'en velocities.
      The dashed line shows the curve $E_{in} \propto c_{\mathcal{A}}^2$, while
      the continuous line shows $E_{in}$ as a function of $c_{\mathcal{A}}$ as obtained
      from equation~(\ref{eq:amp}) for $\alpha = 0$ corresponding to a Kolmogorov
      spectrum.
      The actual growth of $E_{in}$, both simulated and derived from (\ref{eq:amp}),
      show that the growth is less than quadratical but higher than in the simple
      Kolmogorov case.\\[0pt]
     \label{fig:ein}}  
\end{figure}

In our runs, larger values of $c_{\mathcal{A}}$, i.e. of the parameter $f$~(\ref{eq:fsc}),
lead to larger magnetic energy and total energies, while the kinetic energy 
remains smaller than magnetic energy and increases much less
(increasing its value by a factor of $6$ from $c_{\mathcal{A}}=50$ to
$c_{\mathcal{A}}=1000$). In particular Figure~\ref{fig:ein}
shows total energy at the injection scales (see \S~\ref{sec:bc}),
i.e.\ the sum of the modes $n=3$ and $n=4$ (see eq.~\ref{eq:esp})  
of total energy,
\begin{equation}
E_{in} = E_3 + E_4
\end{equation}
as a function of the non-dimensional Alfv\'en velocity $c_{\mathcal{A}}$.
Their growth is less than quadratical in $c_{\mathcal{A}}$, which implies that
the rms of the velocity and magnetic fields at the injection scale
(or equivalently the Els\"asser fields $\delta z_{in}$) grow less
than linearly. Hence as $c_{\mathcal{A}}$ is increased, the ratio
$\chi$, a measure of the relative strength of the nonlinear interactions at the injection
scale,  decreases: at different values of $c_{\mathcal{A}}$ different regimes of 
weak turbulence are therefore realized at the larger scales of the inertial range, as the 
different spectra in Figure~\ref{fig:multisp} confirm.

The presence of a ``double'' inertial range, with a {\it weak-type}
power-law index at larger scales, and a flatter {\it strong-type} power-law index at smaller scales would 
not affect the overall cascade rate, and therefore the scalings of loop heating with loop parameters.
These are set at the larger scales, and are therefore dependent on the cascade rate determined by the 
{\it weak-type} scaling law, for which a physically motivated phenomenological 
derivation is presented below. We stress that the possible existence of a ``double'' inertial range,
surmised here with scaling laws and somewhat tenuous numerical evidence, does not appear to have been
predicted before and requires substantiating evidence from higher numerical resolution simulations
which are planned for the near future.

\subsection{Phenomenology of the Inertial Range and 
Coronal Heating Scalings} \label{sec:sca}

We now introduce a phenomenological model
to determine the energy transfer time-scale  $T_{\lambda}$ and
as a consequence the properties of the cascade. This time-scale,
and therefore the different  spectra which result, can only depend 
on the single non-dimensional quantity defining our system,
namely $f ={\ell_c\, v_{\mathcal A}} / {L\, u_{ph}}$~(\ref{eq:fsc}). The simulations 
show that as this parameter is increased 
the spectra steepen leading to a weakened cascade. 
We revert here to \emph{dimensional quantities} for the scalings, so that we can quantify the
 resulting coronal heating rates. 

The Alfv\'en effect is based on the idea that two counterpropagating
Alfv\'en waves interact only for the time $\tau_{\parallel} = \ell_{\parallel} / v_{\mathcal{A}}$,
leading to a transfer energy time longer that the ``generalized'' eddy turnover time
\begin{equation}
\tau_{\lambda} = \frac{\lambda}{\delta z_{\lambda}},
\end{equation}
The ratio between these two timescales
\begin{equation}
\chi =   \frac{\tau_{\mathcal{A}}}{\tau_{\lambda}}=
 \frac{\ell_{\parallel}\, \delta z_{\lambda}}{\lambda\, c_{\mathcal{A}}}
\end{equation}
gives a measure of their relative strength. 
\cite{iro64} and \cite{kra65} proposed that the energy transfer
time $T_{\lambda}$, because of the Alfv\'en effect,
is longer than the eddy turnover time, and is given by
\begin{equation} \label{eq:ik}
T_{\lambda} \sim \tau_{\lambda} \, \frac{\tau_{\lambda}}{\tau_{A}}, 
\end{equation}
where however they considered  an isotropic situation,
so that the Alfv\'en time was given by the propagation time over the scale 
of the Alfv\'enic packet. For weak turbulence however $\ell_{\parallel} > L$, so that
the Alfv\'en time must be based on the scale $L$:
$\tau_{\mathcal{A}} = L / v_{\mathcal{A}}$.

In addition, we must allow line-tying which acts to slow the
destruction of eddies on a given scale ${\lambda}$ more effectively than the 
standard random encounter effect $ {\tau_{\lambda}}/{\tau_{A}}$ \citep{DMV80}. 
We can therefore  assume a sub-diffusive behaviour for ${\bf z^+} --- {\bf z^-}$ non-linear encounters
leading to 
\begin{equation} \label{eq:gik}
T_{\lambda} \sim \tau_{\lambda} \, 
\left( \frac{\tau_{\lambda}}{\tau_{A}} \right)^{\alpha}, 
\end{equation}
with values $\alpha >1$ and depending in some way on the parameter $f$
[recall that $\alpha=0,1$ correspond respectively to anisotropic Kolmogorov and Kraichnan cases 
(the latter leading to a $k^{-2}$ inertial range spectrum)].

Our simulations then close this ansatz by determining how $alpha$ depends on $f$:
integrating over the whole volume ($\ell \times \ell \times L$), the energy cascade rate
may now be written as
\begin{equation} \label{eq:etr}
\epsilon \sim \ell^2\, L\, \rho\, 
\frac{{\delta z_{\lambda}}^2}{T_{\lambda}},
\end{equation}
Using~(\ref{eq:gik}) the energy transfer rate is given by
\begin{equation} \label{eq:sbe1}
\epsilon \sim \ell^2 L\cdot \rho\, \frac{\delta z_{\lambda}^2}{T_{\lambda}} 
\sim \ell^2 L\cdot \rho\, \left( \frac{L}{v_{A}} \right)^{\alpha} \, 
\frac{\delta z_{\lambda}^{\alpha + 3}}{\lambda^{\alpha + 1}}.
\end{equation}
Identifying, as usual, the eddy energy with the band-integrated Fourier
spectrum $\delta z^2_{\lambda} \sim k_{\perp} E_{k_{\perp}}$, 
where $k_{\perp} \sim \ell / \lambda$,
from eq.~(\ref{eq:sbe1}) we obtain the spectrum
\begin{equation} \label{eq:gsp}
E_{k_{\perp}} \propto k_{\perp}^{ - \frac{3 \alpha + 5}{\alpha + 3} },
\end{equation}
where for $\alpha = 0,1$ the $-5/3,-3/2$ slope for the anisotropic Kolmogorov, Kraichnan
spectra are is recovered, but steeper spectral slopes up to an
asymptotic value of $-3$ are obtained with \emph{higher values of $\alpha$}.

Correspondingly, from eqs.~(\ref{eq:etr})-(\ref{eq:sbe1}), 
the following scaling relations for $\delta z_{\lambda}$ and $T_{\lambda}$ follow:
\begin{equation} \label{eq:zsc}
\delta z_{\lambda} \sim 
\left( \frac{\epsilon}{\ell^2 L \rho} \right)^{\frac{1}{\alpha + 3}}
\left( \frac{v_{\mathcal{A}}}{ L} \right)^{\frac{\alpha}{\alpha + 3}} \
\lambda^{\frac{\alpha + 1}{\alpha + 3}}
\end{equation}
\begin{equation} \label{eq:tsc}
T_{\lambda} \sim 
\left( \frac{\ell^2 L \rho}{\epsilon} \right)^{\frac{\alpha + 1}{\alpha + 3}}
\left( \frac{v_{\mathcal{A}}}{ L} \right)^{\frac{2 \alpha}{\alpha + 3}} \
\lambda^{2\, \frac{\alpha + 1}{\alpha + 3}}
\end{equation}

Recently \cite{bd05} has proposed a similar model,
which aims to overcome some discrepancies between previous models
and numerical simulations, and that self-consistently accounts for the formation
of current sheets, for the cascade of strong 
turbulence.
His energy transfer time is given by
\begin{equation} \label{eq:bd}
T_{\lambda} = \frac{\lambda}{\delta z_{\lambda}} 
\left( \frac{v_{\mathcal{A}}}{\delta z_{\lambda}} \right)^{\alpha},
\end{equation}
but he suggests the interval $ 0 \le \alpha \le 1$ as appropriate to strong turbulence.

As pointed out above of \S~\ref{sec:an},  the solutions of 
equations~(\ref{eq:els1})-(\ref{eq:els3}) depend only on the
non-dimensional parameter $f=\ell_c\, v_{\mathcal{A}} / L\, u_{ph}$
(eq.~(\ref{eq:fsc})) and so $\alpha$ (\ref{eq:gik}) is only a function of $f$
\begin{equation} \label{eq:af}
\alpha = \alpha \left( \frac{\ell_c\, v_{\mathcal{A}}}{L\, u_{ph}} \right).
\end{equation}

We estimate the value of $\alpha$ from the
slope of the total energy spectra~(\ref{eq:gsp}), as described in
\cite{rap07}. As shown in Figure~\ref{fig:multisp} to different
values of $c_{\mathcal{A}} = v_{\mathcal{A}} / u_{ph}$, (i.e.\ $f$) 
ranging from $50$ up to $1000$ correspond spectral slopes
from $\sim -1.8$ up to $\sim -2.7$. Thes in turn
correspond (through eq.~(\ref{eq:gsp})) to values of $\alpha$ 
ranging from  $\sim 0.33$ up to $\sim 10.33$.

How do the above results  affect coronal heating
scalings? The energy that is injected at the large scales by photospheric
motions, and whose energy rate ($\epsilon_{in}$) is given quantitatively by the Poynting
flux~(\ref{eq:tsz}), is transported (without being dissipated) along the inertial
range at the rate $\epsilon$~(\ref{eq:sbe1}), to be finally dissipated 
at the rate $\epsilon_d$.  In a stationary state all these fluxes
must be equal
\begin{equation} \label{eq:bal}
\epsilon_{in}  = \epsilon =  \epsilon_{d}
\end{equation}

The injection energy rate~(\ref{eq:tsz}) is given by
S, the Poynting flux integrated over the photospheric surfaces: 
\begin{multline} \label{eq:dtsz}
\epsilon_{in} = S = \\
= \rho v_{\mathcal A} 
\left[ \int_{z=L} \! \mathrm{d} a\, \left( \boldsymbol{u}_{\perp}^L 
\cdot \boldsymbol{b}_{\perp}  \right)
-  \int_{z=0} \! \mathrm{d} a\, \left( \boldsymbol{u}_{\perp}^0 
\cdot \boldsymbol{b}_{\perp}  \right) \right].
\end{multline}
2D spatial periodicity in the orthogonal
planes allows us to expand the velocity and magnetic fields in Fourier series, e.g.\
\begin{equation} \label{eq:2dfe}
\mathbf{u}_{\perp} \left( x, y \right) = \sum_{r,s} \mathbf{u}_{r,s}\, e^{i \mathbf{k}_{r,s} \cdot \mathbf{x}},
\end{equation}
where
\begin{equation}
\mathbf{k}_{r,s} = \frac{2\pi}{\ell} \left( r, s, 0 \right)
\qquad r, s \in \mathbb{Z}
\end{equation}
The surface integrated scalar product of $\mathbf{u}_{\perp}$
and $\mathbf{b}_{\perp}$  at the boundary is then given by
\begin{multline} \label{eq:par}
\int \! \mathrm{d} a\, \boldsymbol{u}_{\perp}
\cdot \boldsymbol{b}_{\perp} 
= \sum_{r,s} \boldsymbol{u}_{r,s}  \cdot 
\int_0^{\ell} \! \! \! \int_0^{\ell} \mathrm{d}x  \mathrm{d}y \
\boldsymbol{b}_{\perp} e^{i \boldsymbol{k}_{r,s} \cdot \boldsymbol{x}} = \\
= \ell^2\,  \sum_{r,s}  \boldsymbol{u}_{r,s} \cdot \boldsymbol{b}_{-r,-s}, 
\qquad  r, s \in \mathbb{Z}
\end{multline}
This integral is clearly dominated by large scales consistent with observations
of photospheric motions.
In our case (eq.~(\ref{eq:bc})) boundary velocities only have components
for wave numbers $(r, s) \in \mathbb{Z}^2$  with absolute values between $3$ and $4$,
$3 \le  \left( r^2 + s^2 \right)^{1/2} \le 4$. Then in~(\ref{eq:par}) only the 
corresponding components of $\mathbf{b}_{\perp}$ are selected.

At the injection scale, which is the scale of convective motions
$\ell_c  \sim 1,000\, km$, a weak 
turbulence regime develops, so that the cascade 
along the axial direction $z$ is limited and in particular the magnetic field 
$\mathbf{b}_{\perp}$ can be considered approximately uniform along
$z$ at the large orthogonal scales. Then from eq.~(\ref{eq:dtsz}) we obtain
\begin{equation} \label{eq:dein}
\epsilon_{in} = S \sim \rho v_{\mathcal A}
\int \! \mathrm{d} a\, \left( \mathbf{u}_{\perp}^L - \mathbf{u}_{\perp}^0 \right)
\cdot \mathbf{b}_{\perp}
\end{equation}
Introducing 
$\mathbf{u}_{ph} = \mathbf{u}_{\perp}^L - \mathbf{u}_{\perp}^0$, 
using relation~(\ref{eq:par}), and integrating over the surface,  we can now write
\begin{equation} \label{eq:dein2}
\epsilon_{in} = S \sim  \ell^2 \rho v_{\mathcal A}
u_{ph}  \delta z_{\ell_c},
\end{equation}
where we have approximated the value of $\delta b_{\ell_c}$, the rms of the magnetic
field at the injection scale $\ell_c$, with the rms of the Els\"asser variable
because the system is magnetically dominated, i.e.\ 
$\delta z_{\ell_c} = \left( \delta u^2_{\ell_c} + \delta b^2_{\ell_c} \right)^{1/2} \sim \delta b_{\ell_c}$.

We now have an expression for $\epsilon_{in}$, where the only unknown variable is
$\delta z_{\ell_c}$, as $\ell_c$, $\rho$, $v_{\mathcal{A}}$ and $u_{ph}$ are the parameters
characterizing our model of a coronal loop.

The transfer energy rate $\epsilon$ does not depend on $\lambda$. 
Considering then $\lambda=\ell_c$
in equation~(\ref{eq:sbe1}), we have
\begin{equation} \label{eq:sbe2}
\epsilon 
\sim \frac{\rho\, \ell^2 L^{\alpha+1}}{\ell_c^{\alpha+1} \, v_{A}^{\alpha}} \, 
\delta z_{\ell_c}^{\alpha + 3}.
\end{equation}

Equations~(\ref{eq:dein2}) and (\ref{eq:sbe2}) show another aspect of self-organization.
Both $\epsilon_{in}$ and $\epsilon$, respectively the rate of the energy flowing in 
the system at the large scales, and the rate of the energy flowing from the large 
scales toward the small scales depend on $\delta z_{\ell_c}$, the rms of the fields at 
the large scale.
This shows that the energetic balance of the system is determined by the balance
of the energy fluxes  $\epsilon$ and $\epsilon_{in}$ at the large scales.
The small scales will then dissipate the energy that is transported along the inertial range
(see eq.~(\ref{eq:bal})). This implies that \emph{beyond a numerical threshold total
dissipation (dissipation integrated over the whole volume)
is independent of the Reynolds number}. In fact beyond a value of 
the Reynolds number for which the diffusive time at the large scale is negligible,
i.e.\ when the resolution is high enough to resolve an inertial range, the large-scale
balance between $\epsilon$ and $\epsilon_{in}$ is no longer influenced by diffusive 
processes. Of course this threshold is quite low respect to the high values
of the Reynolds numbers for the solar corona, but it is still computationally very demanding. 

An analytical expression for the coronal heating scalings may be obtained
from (\ref{eq:dein2}) and (\ref{eq:sbe2}) yielding the value of $\delta z_{\ell_c}^{\ast}$
for which the balance $\epsilon_{in} = \epsilon$ is realized: 
\begin{equation} \label{eq:amp}
\frac{\delta z_{\ell_c}^{\ast}}{u_{ph}} 
\sim \left( \frac{\ell_c v_{A}}{L u_{ph}} \right)^{\frac{\alpha + 1}{\alpha + 2}}
\end{equation}
Substituting this value in (\ref{eq:sbe2}) or equivalently in (\ref{eq:dein2}) we obtain
the energy flux
\begin{equation} \label{eq:chs}
S^{\ast} 
\sim \ell^2 \rho \, v_{A} u_{ph}^2 
\left( \frac{\ell_c v_{A}}{L u_{ph}} \right)^{\frac{\alpha+1}{\alpha+2}}.
\end{equation}
As stated in~(\ref{eq:bal}) in a stationary cascade all energy fluxes are equal on the average.
$S^{\ast}$ is then the energy that for unitary time flows through the boundaries in the
coronal loop at the convection cell scale, and that from these scales flows towards 
the small scales.
This is also the dissipation rate, and hence the \emph{coronal heating scaling,
i.e.\ the energy which is dissipated in the whole volume for unitary time.}
As shown in equation~(\ref{eq:af}) the power $\alpha$ depends on the parameters
of the coronal loop, and its value is determined numerically with the 
aforementioned technique.

\begin{figure}
     \includegraphics[width=0.47\textwidth]{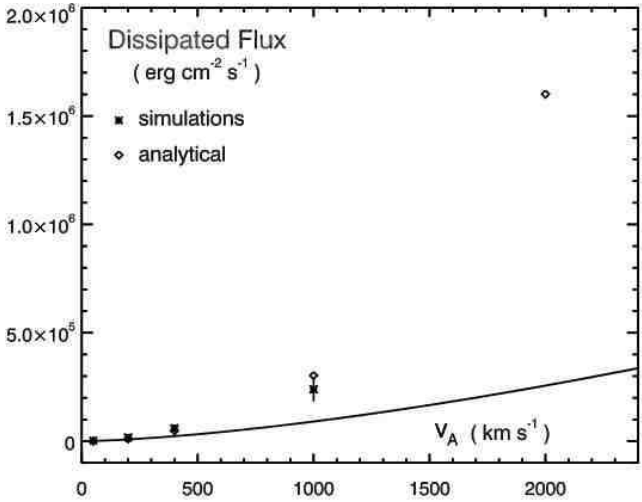}
      \caption{Analytical (\ref{eq:fchs}) and numerically computed dissipated 
               flux as a function of the axial Alfv\'en velocity $v_{\mathcal{A}}$. 
               The continuous line shows the Poynting flux (\ref{eq:fchs}) as a function
               of  $v_{\mathcal{A}}$ in the case $\alpha = 0$, corresponding to 
               a Kolmogorov-like cascade. To higher values of  $v_{\mathcal{A}}$
               correspond a higher dissipation rate because a weak turbulence
               regime develops.\\[0pt]
     \label{fig:dissA}}  
\end{figure}

The observational constraint with which to compare our results is 
the energy flux sustaining an active region.
The energy flux at the boundary is the axial component of the Poynting 
vector $S_z$ (see \S~\ref{sec:ee}).
This is obtained dividing $S^{\ast}$~(\ref{eq:chs}), the Poynting flux
integrated over the surface, by the surface $\ell^2$:
\begin{equation} \label{eq:fchs}
S_z = \frac{S^{\ast}}{\ell^2} \sim \rho \, v_{A} u_{ph}^2 
\left( \frac{\ell_c v_{A}}{L u_{ph}} \right)^{\frac{\alpha+1}{\alpha+2}},
\end{equation}
where $\alpha$ is not a constant, but a function of the 
loop parameters (\ref{eq:af}).
The exponent in (\ref{eq:fchs}) goes from $0.5$ for $\alpha = 0$
up to the asymptotic value $1$ for larger $\alpha$.
We determine $\alpha$ numerically, measuring the slope of the
inertial range (Figure~\ref{fig:multisp}), and inverting the
spectral power index~(\ref{eq:gsp}). 
We have used simulations
F, G, H and I to compute the values of $\alpha$, because they
implement hyperdiffusion, resolve the inertial range, and then
are beyond the numerical threshold below which total dissipation
does not depend on the Reynolds number. These simulations implement
$v_{\mathcal{A}} = 50$, $200$, $400$ and $1,000$, 
and the corresponding $\alpha$ are $\sim 0.33$, $1$, $3$, $10.33$.
The corresponding values for the power $(\alpha+1)/(\alpha+2)$ (\ref{eq:fchs})
are $\sim 0.58$, $0.67$, $0.8$ and $0.91$, close to the asymptotic value $1$.
$S_z$ is shown in 
Figure~\ref{fig:dissA} (diamond points) as a function of the axial 
Alfv\'en velocity $v_{\mathcal{A}}$. 
To compute the value of $S_z$ for $v_{\mathcal{A}} = 2,000\, km\, s^{-1}$
we have estimated $\alpha \sim .95$, although for values close to $1$
$S_z$ does not have a critical dependence on the value of the exponent.

In Figure~\ref{fig:dissA} we compare the analytical function 
$S_z$~(\ref{eq:fchs}) with the respective value determined
from our numerical simulations (star points), i.e.\ with the total 
dissipation rate by the surface and converted to dimensional units 
($(J+\Omega)/\ell^2$, see (\ref{eq:tdiss})). 
For the numerical simulation values, the error-bar is defined as 1 
standard deviation of the temporal signal.
The analytical and computational values are in good agreement
for all the 4 simulations considered, and for the more realistical
value $v_{\mathcal{A}} = 2,000\, km\, s^{-1}$ the dissipated flux
is $\sim 1.6\times10^6\, erg\, cm^{-2}\, s^{-1}$. This value is in the
lower range of the observed constraint $10^7\, erg\, cm^{-2}\, s^{-1}$.

The continuous line in Figure~\ref{fig:dissA} corresponds to the 
function $S_z$ for $\alpha = 0$ (which is approximately realized for 
$v_{\mathcal{A}} \lesssim 50\, km\, s^{-1}$), in 
correspondence of which a Kolmogorov spectrum would be present,
and $S_z \propto v_{\mathcal{A}}^{3/2}$. The computed and analytical
values of $S_z$ for higher $v_{\mathcal{A}}$ are always beyond this curve,
because $\alpha$ increases its values, and a more
efficient dissipation takes place. This is due to the fact that to higher values
of $\alpha$ correspond higher values of the energy transfer time, 
and consequently a longer linear stage, higher values of the fields at the large 
scales~(\ref{eq:amp}), and hence a higher value of the energy rates
(see (\ref{eq:dein2}), (\ref{eq:sbe2}) and (\ref{eq:chs})).
So that it is realized the only apparently paradox that to a weaker turbulent regime, 
to which corresponds less efficiency in the nonlinear terms, corresponds
a higher total dissipation.

In the last paragraph of \S~\ref{sec:ee} we have shown that when the condition~(\ref{eq:efc})
is satisfied the emerging flux can be neglected. But in eq.~(\ref{eq:efc}) we have to specify
the value of the magnetic field $b_{\perp}^{turb}$ self-consistently generated by the non-linear
dynamics. This value is given by (\ref{eq:amp}) as the magnetic field dominates
($\delta z^{\ast}_{\ell_c} \sim b_{\perp}^{turb}$). By substitution we can now estimate
that the emerging flux is negligible when the emerging component of the magnetic field
satisfies
\begin{equation} \label{eq:efcc}
b_{\perp}^{ef} < B_0\, \sqrt{ 
\left( \frac{\ell_c}{L} \right)^{\frac{\alpha+1}{\alpha+2}}
\left( \frac{u_{ph}}{v_{\mathcal{A}}} \right)^{\frac{1}{\alpha+2}}
}
\end{equation}
In the asymptotic state $\alpha \gg 1$ the condition reduces to 
$b_{\perp}^{ef} / B_0 <  \sqrt{ \ell_c / L }$. For a coronal loop
with $L \sim 40,000\, km$, as $\ell_c \sim 1,000\, km$ this implies
that emerging flux does not play a role if $b_{\perp}^{ef} / B_0 < 1/ 6$.

\begin{figure}
     \includegraphics[width=0.47\textwidth]{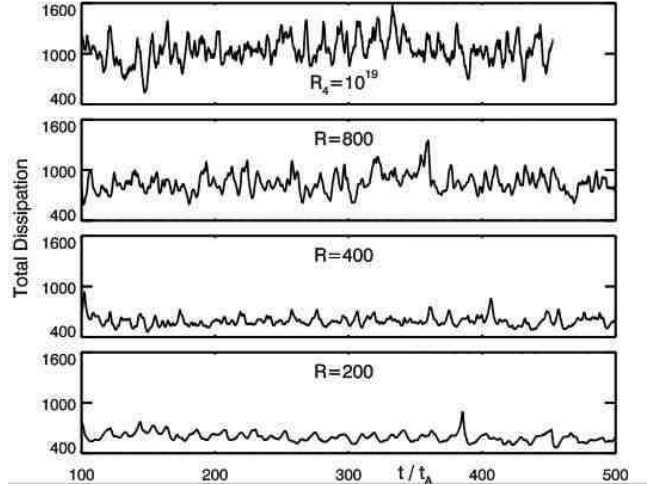}
      \caption{\emph{Transition to turbulence}: Total ohmic and viscous dissipation 
               as a function of time for simulations~A, B, C, G (displayed on 
               the same scales).
               All the simulations implement $c_{\mathcal{A}} = 200$, but different
               Reynolds numbers, from $Re=200$ up to $800$. Run~G implements
               hyperdiffusion. For Reynolds numbers lower than $100$ the signal
               is completely flat and displays no dynamics, at higher Reynolds
               smaller temporal structures are present.\\[0pt]
     \label{fig:turbod}}  
\end{figure}

\subsection{Transition to Turbulence and 
Dissipation vs.\ Reynolds Number} \label{sec:trans}

Turbulence is a characteristic of high Reynolds number systems 
(e.g.\ \cite{fr95}).
For a sufficiently high viscosity nonlinear dynamics is strongly
suppressed, and our system relaxes to a diffusive equilibrium
(\S\ref{sec:eod}), and no significant small scale is formed.
Increasing the Reynolds number, the diffusive time at the
injection scale~(\ref{eq:taud}) $\tau_d \sim Re\, \ell_c^2$
increases. At a certain point it will be big enough
not to influence the dynamics as the large scales,
an inertial range will then be resolved and
total dissipation will not depend any longer from the Reynolds numbers.
In fact for higher values of $Re$ the intertial range will extend
to higher wave-numbers, but the energy flux will remain the same.

\begin{figure}
     \includegraphics[width=0.47\textwidth]{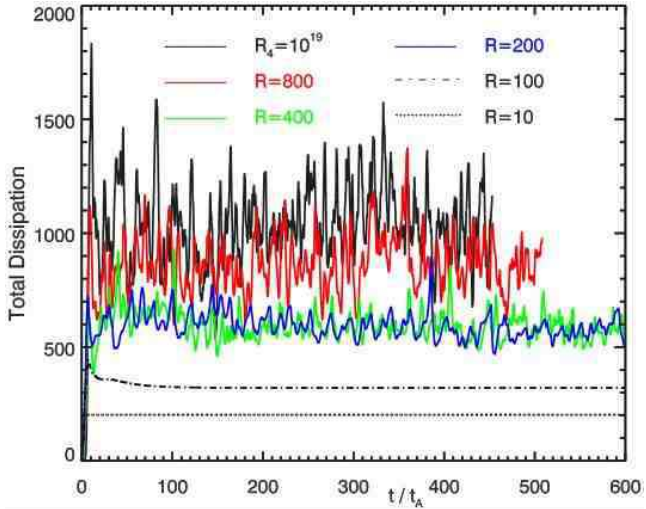}
      \caption{Total ohmic and viscous dissipation 
               as a function of time for simulations~A, B, C, D, E and 
               G, all of them implement $c_{\mathcal{A}} = 200$ but different
               Reynolds numbers. The threshold beyond which dissipation
               is independent of the Reynolds number can be identified
               around $Re = 800$, corresponding to a numerical resolution
               of 512x512 points in the orthogonal planes.\\[0pt]
     \label{fig:multid}}  
\end{figure}

At higher Reynolds numbers smaller scales are resolved, and 
each scale will contribute with its characteristic time
$T_{\lambda}$ to the temporal structure of the rms of the
system. Figure~\ref{fig:turbod} shows total dissipation
as a function of time for simulations A, B, C and G, 
on the same time interval, and on the same scale.
At increasingly higher values of the Reynolds numbers smaller and smaller
temporal structures are added to the signal. 
Ideally the temporal structure of total dissipation
at higher Reynolds numbers is well described by 
shell-model simulations.
For smaller values of Re the signal is completely flat (see 
Figure~\ref{fig:multid}). 
This behaviour identifies a \emph{transition to turbulence}.

Figure~\ref{fig:multid} shows total dissipation as a function
of time for the same 4 simulations shown in Figure~\ref{fig:turbod},
plus other 2 simulations at lower Reynolds number, respectively
$Re =100$ and $Re=10$ for the complete time interval.
For the lowest value of $Re$ no dynamics is present, so that
the threshold value for the transition to turbulence can 
be set to $Re \sim 100$.
To higher values of $Re$ dissipation grows. An inertial range
is barely solved with a resolution of 512x512 grid points
in the $x$-$y$ planes, so that simulation with $Re = 800$ can 
be considered at the threshold.
On the other hand simulation~G implements hyperdiffusion, 
so that an inertial range is solved, and the dynamics is not
affected by diffusion. The presence of  a sufficiently extended inertial
range implies in fact that we are beyond the numerical threshold where
dissipation does not depend on the Reynolds number (\S~\ref{sec:sca}). 
The threshold value can be identified to a sufficient extent
at $Re =800$, i.e.\ for a numerical grid of 512x512 points.
The number of points to use along the axial direction should
be enough to allow the formation of all the small scales due
to the ``critical balance'' (Figure~\ref{fig:spzp}), but a larger number of points would
only result in a waste of computational time.

\subsection{Inverse Cascade and Line-tying} \label{sec:lt}

Two dimensional simulations \citep{ein96,georg98} have shown an 
inverse cascade for the magnetic energy, corresponding in physical 
space to the coalescence of magnetic islands.
In the 3D case the DC magnetic field along the axial direction
is present, giving rise to a field line tension that tends to
inhibit an inverse cascade, as motions linked to the coalescence would
bend the field lines of the total magnetic field, which are mostly elongated 
along the axial direction (Figure~18). 
On the other hand field line tension depends on the strength of the axial field,
becoming stronger for a stronger field.

\begin{figure}
     \includegraphics[width=0.47\textwidth]{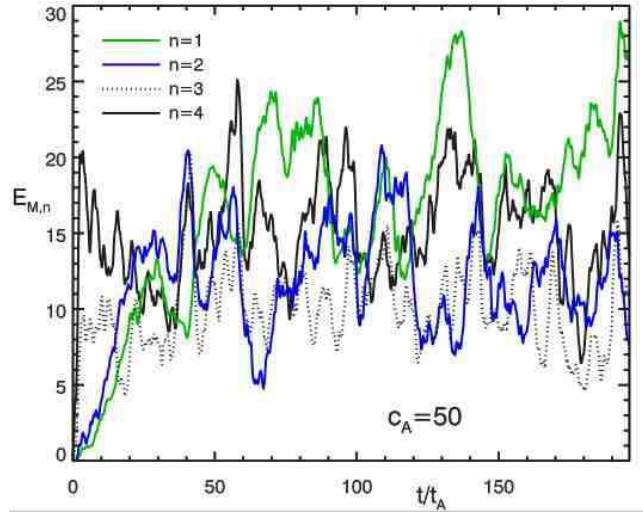}
      \caption{\emph{Run F:} In this simulation with $c_{\mathcal{A}} = 50$
               an inverse cascade at the wavenumbers $n=1$ and $2$ is realized.
               Energy is injected at wavenumbers $n=3$ and $4$.\\[0pt]
     \label{fig:fig14}}  
\end{figure}

In Figures~\ref{fig:fig14} and \ref{fig:fig15} the first 4 modes of magnetic
energy for simulations~F and I, respectively with $c_{\mathcal{A}} = 50$ and
$1000$, are plotted as a function of time.
Energy is injected at wave-numbers $n=3$ and $4$.
Modes associated to  wave-numbers 1 and 2 grow to higher values than at the injection scale
in run~F, while in run~I they are always limited to lower values.
In runs G and H, with respectively $c_{\mathcal{A}} = 200$ and $400$
an intermediate behaviour is found, but none of the modes $n=1$ or $2$
never becomes bigger than the injection energy modes.

\subsection{Timescales} \label{sec:tmsc}

In the previous sections we have always affirmed that the Alfv\'en crossing
time $\tau_{\mathcal{A}} = L / v_{\mathcal{A}}$ is the fastest timescale in
the system, and that in particular it is smaller than the nonlinear 
timescale $\tau_{nl}$, which we can identify with the energy transfer
time (\ref{eq:tsc}) at the injection scale $\tau_{nl} = T_{\ell_c}$.

In Figure~\ref{fig:fig3} it is already clear that the nonlinear timescale
is longer that $\tau_{\mathcal{A}}$, in fact it shows
that the timescale over which energy has substantial variations is
bigger than the Alfv\'en crossing time.

\begin{figure}
     \includegraphics[width=0.47\textwidth]{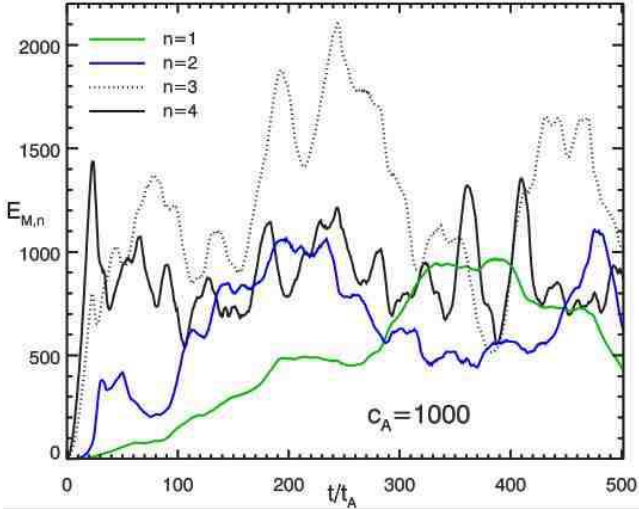}
      \caption{\emph{Run I:} Simulation performed with $c_{\mathcal{A}} = 1000$.
               The increased magnetic field line tension inhibits an inverse
               cascade for the orthogonal magnetic field.\\[0pt]
     \label{fig:fig15}}  
\end{figure}

The same behaviour is identified in
Figures~\ref{fig:fig14}-\ref{fig:fig15}, which show the time evolution of
the magnetic energy modes for runs~F and I. These are more relevant
quantities, because to realize a weak MHD turbulence regime it is 
required that the energy transfer time $T_{\lambda}$ is bigger than
the crossing time $\tau_{\mathcal{A}}$ at the injection scale
$\lambda = \ell_c$ and for a limited range of smaller scales down
to some lower bound $\lambda^{\ast}$:
$\lambda^{\ast} \le \lambda \le \ell_c$. The magnetic energy modes
at the injection scale ($n=3$ and $4$) change their values on scales
bigger than $\tau_{\mathcal{A}}$, and for a larger value of the
Alfv\'en velocity the nonlinear timescale is longer respect to the
crossing time (Figures~\ref{fig:fig14}-\ref{fig:fig15}).
We can roughly estimate $\tau_{nl} \sim 5 \tau_{\mathcal{A}}$ for run~F
with $c_{\mathcal{A}} = 50$ and $\tau_{nl} \sim 20 \tau_{\mathcal{A}}$
for run~I with $c_{\mathcal{A}} = 1,000$.

Using our scaling relations we can derive an analytical estimate for
the energy transfer time $T_{\lambda}$. Substituting the energy rate
(\ref{eq:chs}) in equation~(\ref{eq:tsc}) we obtain:
\begin{equation} \label{eq:tsc1}
T_{\lambda} \sim 
\left( \tau_{\mathcal{A}} \tau_c^{\alpha+1} \right)^{\frac{1}{\alpha+2}}
\left( \frac{\lambda}{\ell_c} \right)^{2 \frac{\alpha+1}{\alpha+3}},
\end{equation}
where $\tau_c = \ell_c / u_{ph}$. In particular the ratio over
the Alfv\'en crossing time is:
\begin{equation} \label{eq:tsc2}
\frac{T_{\lambda}}{\tau_{\mathcal{A}}} \sim 
\left( \frac{\tau_c}{\tau_{\mathcal{A}}} \right)^{\frac{\alpha+1}{\alpha+2}}
\left( \frac{\lambda}{\ell_c} \right)^{2 \frac{\alpha+1}{\alpha+3}},
\end{equation}
and as $\tau_c > \tau_{\mathcal{A}}$ then self-consistently 
$T_{\lambda} > \tau_{\mathcal{A}}$.
For our loop $\ell_c \sim 1,000\, km$ and $u_{ph} \sim 1\, km\, s^{-1}$,
so that $\tau_c \sim 1,000\, s$. For runs~F and I shown in 
Figures~\ref{fig:fig14} and \ref{fig:fig15},
the loop length is always $L = 40,000\, km$, while the Alfv\'en velocity
is respectively $v_{\mathcal{A}} = 50$
and $1,000\, km\, s^{-1}$, and the corresponding crossing times
$\tau_{\mathcal{A}} = 800$ and $40\, s$.
Using the values of $\alpha$ computed in \S\ref{sec:sca}
(respectively $\alpha = 0.33$ and $10.33$)
we can then roughly estimate from (\ref{eq:tsc2}), the nonlinear
timescale $\tau_{nl}=T_{\lambda = \ell_c}$ and its ratio with the 
Alfv\'en crossing time:
\begin{equation} \label{eq:tsc3}
\frac{\tau_{nl}}{\tau_{\mathcal{A}}} =
\frac{T_{\ell_c}}{\tau_{\mathcal{A}}} \sim 
\left( \frac{\tau_c}{\tau_{\mathcal{A}}} \right)^{\frac{\alpha+1}{\alpha+2}}.
\end{equation}
For runs F and I we find $\tau_{nl} / \tau_{\mathcal{A}} = 1.2$ and $22.3$
in agreement with the simulations.

Equation~(\ref{eq:tsc2}) can also be used to estimate the extension of the
weak turbulence inertial range. The region for which the weak turbulence 
condition $T_{\lambda} > \tau_{\mathcal{A}}$ is satisfied is:
\begin{equation}
\lambda > \lambda^{\ast} = 
\ell_c \left( \frac{\tau_{\mathcal{A}}}{\tau_c} \right)^{\frac{\alpha+3}{2(\alpha+2)}}
\end{equation}

\begin{figure}
     \includegraphics[width=0.47\textwidth]{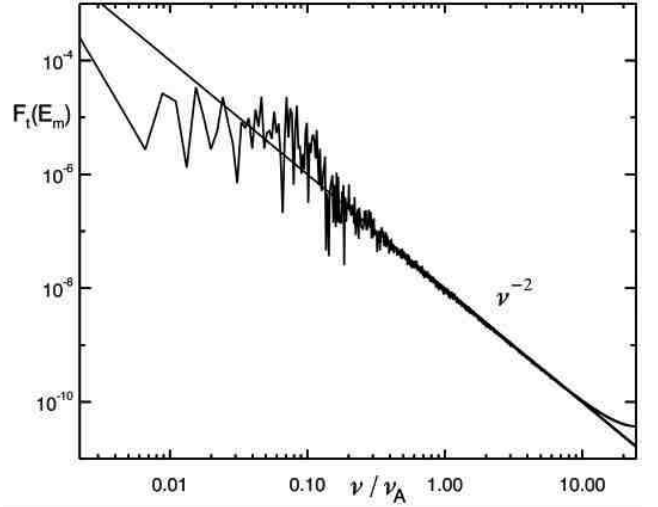}
      \caption{Temporal spectrum of magnetic energy for run~G.
      $\nu_{\mathcal{A}} = 1 / \tau_{\mathcal{A}}$ is the frequency corresponding
      to the Alfv\'en crossing time. The intermediate part of the spectrum exhibits
      a $\nu^{-2}$ power law.\\[0pt]
     \label{fig:fig16}}  
\end{figure}

Figure~\ref{fig:fig16} shows the temporal spectrum of magnetic energy
for run~G with $c_{\mathcal{A}} = 200$, i.e.\ we perform the Fourier
transform of the magnetic energy as a function of time, and then
plot its squared modulus. We use run~G because it is the one for which
we have saved more frequently the rms quantity and then the plot covers
a wider range at high frequencies. The power spectrum is roughly constant
up to $\nu / \nu_{\mathcal{A}} \sim 0.2$, which corresponds to 
$t / \tau_{\mathcal{A}} \sim 5$ in agreement with our scaling~(\ref{eq:tsc3})
which for this case gives $\tau_{nl} / \tau_{\mathcal{A}} \sim 3.3$.
Beyond this critical point the power spectrum exhibits a power law 
which fits $\nu^{-2}$, in agreement with shell-model simulations \citep{buc07}.

\section{DISCUSSION AND CONCLUSIONS} \label{sec:disc}

We would like first to clarify a few concepts that might otherwise result in
misunderstandings of the work that we have presented.
The concept of turbulence is used to describe different processes in different
research fields, so that its use, without specifications, can result
vague and misleading. It is in fact very often used to describe 
chaotic behaviors at the small scales, often linked to the intermittent
dissipation of energy.  Although this aspect is present 
in our simulations, when we say that the Parker problem  is an MHD turbulence problem, 
we refer mainly to the property of turbulence to transfer energy from large to 
small scales.  Namely to its ability to transport the energy from the scale of 
photospheric motions ($\sim 1000\, km$), where it is injected, down to the small 
dissipative scales (meters?), without dissipating it at the intermediate scales. This property is clearly identified
by the presence of an inertial range with a power law spectrum, which extends
from the injection scale to the dissipative scale.

Furthermore turbulence, magnetic reconnection and ohmic heating associated to currents
are sometime presented as alternative and/or mutually exclusive coronal heating models.
This contraposition is artificial. Current 
sheets are in fact {\bf the dissipative structures} of MHD turbulence, and magnetic 
reconnection at the loci of  current sheets is observed in virtually every MHD turbulence 
simulation in both 2D and 3D (see e.g., \cite{bisk03} and references
therein ). Nanoflares are then naturally associated with the time and space intermittency of the 
small scale deposition of energy (as shown in the 2D case by \cite{georg98}), which is due
to the cascade which leads to the formation and dissipation of current sheets,  and to which we refer
collectively with the term MHD turbulence.

In summary, the main results presented in this paper are the following:
\begin{itemize}
\item The time-dependent Parker problem may be seen as an
MHD turbulence problem, where the large scale forcing function is realized 
by the photospheric motions.
\item This system is genuinely turbulent, in the sense that small scale formation is not driven passively
by the random walk of the footpoints, rather it is a property of the maxwell stresses developing in the coronal volume.
Current sheets therefore do not generally result \emph{directly} from
a ``geometrical'' misalignment of neighboring magnetic field lines
stirred by their footpoint (random) motions, \emph{they are the
result of a nonlinear cascade in a self-organized system}.
\item Nanoflares are naturally associated with the intermittent dissipation of the 
energy that, injected at the
large scales by photospheric motions, is transported to the dissipative
scales through a cascade, and is finally dissipated through nonlinear
magnetic reconnection. 
\item Beyond a threshold, that is low compared to the coronal Reynolds numbers,
but still computationally very demanding, total dissipation is independent
of the Reynolds numbers. This threshold corresponds to a numerical resolution
of $\sim 512\times512$ grid points in the planes orthogonal to the dominant
DC magnetic field.
\item As the loop parameters vary, different regimes of
turbulence develop: strong turbulence is found for weak axial magnetic fields and long loops, leading to Kolmogorov-like
spectra in the perpendicular direction, while weaker and weaker regimes (steeper spectral slopes of total energy)
are found for strong axial magnetic fields and short loops. 
There is no single universal scaling law (see (\ref{eq:fchs})), 
as a consequence  the scaling of the heating rate with axial magnetic field intensity, which depends on the spectral index of total energy for given loop parameters, must vary from 
$B_0^{3/2}$ for weak fields to $B_0^2$ for strong fields at a given aspect ratio.
 
\item For a loop $40,000\, km$ long , with an Alfv\'en velocity 
$v_{\mathcal{A}} = 2,000\, km\, s^{-1}$ and a numerical density of 
$10^{10}\, cm^{-3}$, whose footpoints are subject to 
photospheric motions of $u_{ph} \sim 1\, km\, s^{-1}$ on a scale of 
$\ell_c \sim 1,000\, km$,  the energy flux entering the system and being dissipated
is $S_z \sim 1.6\times10^6\, erg\, cm^{-2}\, s^{-1}$. On the other hand, for a
coronal loop typical of a quiet-Sun region, that has the same parameter of the previous
case but with a  length of $100,000\, km$ and $v_{\mathcal{A}} = 500\, km\, s^{-1}$,
the resulting Poynting flux is $S_z \sim 7\times10^4\, erg\, cm^{-2}\, s^{-1}$.
\end{itemize}

The most advanced EUV and X-RAY imagers (e.g.\ those onboard SOHO, TRACE, STEREO 
and HINODE) have space resolutions ($\sim 800\, km$) of the order of the granulation cells.
Hence they do not resolve the small-scales where current sheets,
magnetic reconnection and all the dynamical features of the system
take place. Their resolution is roughly $1/5$ the length of the perpendicular cross-section 
of our numerical box ($\sim 4000\, km$). Hence, even if the system is highly dynamical
on small-scales (see Figure~17 and the associated movie), 
integrating over these scales has the effect to
``averaging'' the small scale dynamics. In particular small 
scale reconnection cannot be detected, magnetic fieldlines will 
appear only slightly bended (Figure~18),
and their dynamics will appear slower (a modulation of the nonlinear timescale
with the thermodinamical timescales). 

The topological and dynamical effects associated with magnetic reconnection should be 
taken into account when modeling the thermodynamical and observational
properties of coronal loops \citep{schr07}, recalling that most of the dynamics take place at sub-resolution scales while 
we observe the integrated emission.

Two density current fields that have the same ``steady'' integrated ohmic dissipation,
balanced by a corresponding Poynting Flux (see \S~\ref{sec:eod}, equation~(\ref{eq:tdiss}) and Figure~\ref{fig:diss}), 
but with different spatial distributions will have different emissions.
Consider the first with only large scale components, as the one that would result
from a diffusive process (\S~\ref{sec:eod}), while
in the second the current has only small scale components,
as in the simulations that we have presented.
In the second case the filling factor is small 
(Figures~17 and 18)
so that the density of current has a far larger value, and this would
correspond to two very different
thermodynamical and observational outcomes. But the highly dynamical
effects associated with the second case will be averaged
and result less dynamical when integrated. Still the integrated observables
should be very distinct between the two cases. 

Finally, while our simulations give an accurate description of the time-dependent Parker problem, with the limitations on the photospheric forcing field described in the introduction,
the use of the reduced MHD equation is justified only
for slender loops threaded by a strong axial magnetic field. For short loops,
or loops that have orthogonal component of the magnetic field comparable
to the axial component, the full set of MHD equation should be implemented.
For the slender loops that we have simulated we observe a modest accumulation of energy,
which subsequently is released via nanoflares. On the other hand shorter loops, 
or loops in a more complicated geometry, or subject to loop-loop interactions,
and more generally loops affected by the neighboring coronal environment, might
exhibit the ability to accumulate more energy (e.g.\ \cite{low06}) and then release it in
larger flares, possibly via a ``secondary instability'' \citep{dahl05} or fast magnetic
reconnection \citep{cas06}. 

Magnetohydrodynamics (MHD) has proved to be a useful tool to investigate the
properties of the turbulent cascade \citep{bisk03}. MHD is very well
known to give an approximate description of the plasma dynamics at \emph{large
scales} and \emph{low frequencies}. In MHD turbulence it is generally supposed
that at the small scales a ``dissipative mechanism'' is present. Most of the properties of the turbulent
cascade do not depend on the details of the dissipative mechanism, whether it
is described by the diffusive operator present in 
equations~(\ref{eq:adim1})-(\ref{eq:adim2}), or 
more properly by a kinetic mechanism.

In particular in our case, the timescales associated at the scale $\lambda$
((\ref{eq:tsc1}) for weak turbulence and (\ref{eq:bd}) for the strong case)
decrease for smaller scales. In this way the small-scale dynamics is
characterized by high-frequency phenomena, and then it is not well
described by MHD, but rather a kinetic model would be more
appropriate. It is then possible that (self-consistently) at the small scales \emph{particle
acceleration} plays an important role in the dissipation of energy, a
physical process that should be investigated through kinetic models.
Nevertheless the coronal heating rates~(\ref{eq:fchs}), like the cascade properties
over an extended range of scales,  are independent 
of the details of the dissipation mechanism. They are determined
by the balance, at the \emph{large scales} (see \S~\ref{sec:sca}), between the rate 
of the energy flowing into the loop from the boundaries due to the work done by 
photospheric motions on the magnetic field line footpoints at the scale of the convective cells, 
and the rate at which the energy flows along the inertial range from the large
scales towards the small scales.

\acknowledgements
The authors would like to thank Bill Matthaeus and the anonymous referee for 
very useful comments.
A.F.R.\ is supported by the NASA Postdoctoral Program, M.V. is supported
by NASA LWS-TR\&T and SR\&T, and R.B.D. is supported by NASA SPTP.
A.F.R. and M.V. thank the IPAM program ``Grand Challenge Problems in 
Computational Astrophysics'' at UCLA. Simulations were carried out on JPL supercomputers.
\\

\begin{figure}[p]
      \centering
      \subfloat[]{
               \label{fig:atsjfl:a}             
               \includegraphics[width=0.46\linewidth]{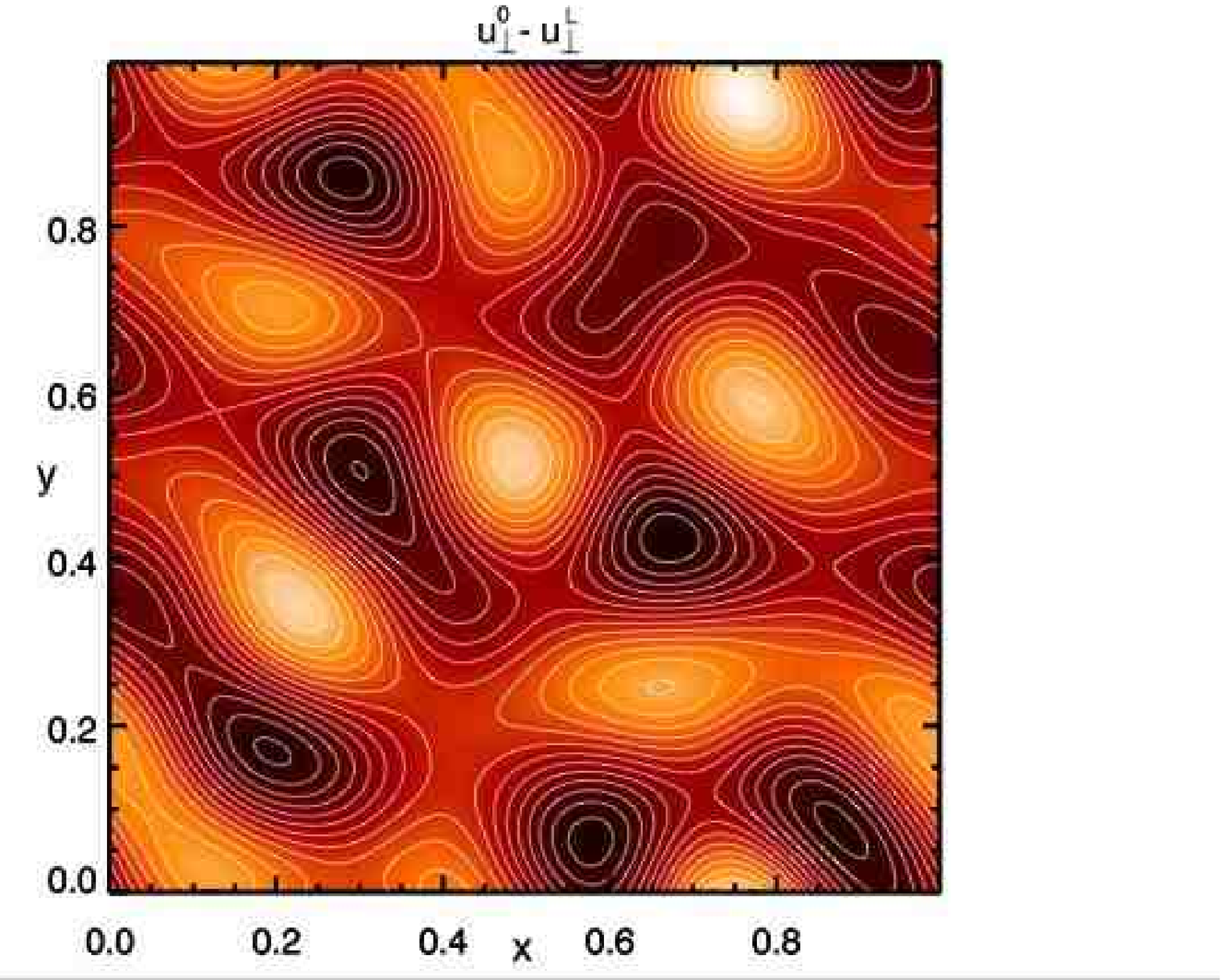}
               }
      \hspace{0.01\linewidth}
      \subfloat[]{
               \label{fig:atsjfl:b}             
               \includegraphics[width=0.46\linewidth]{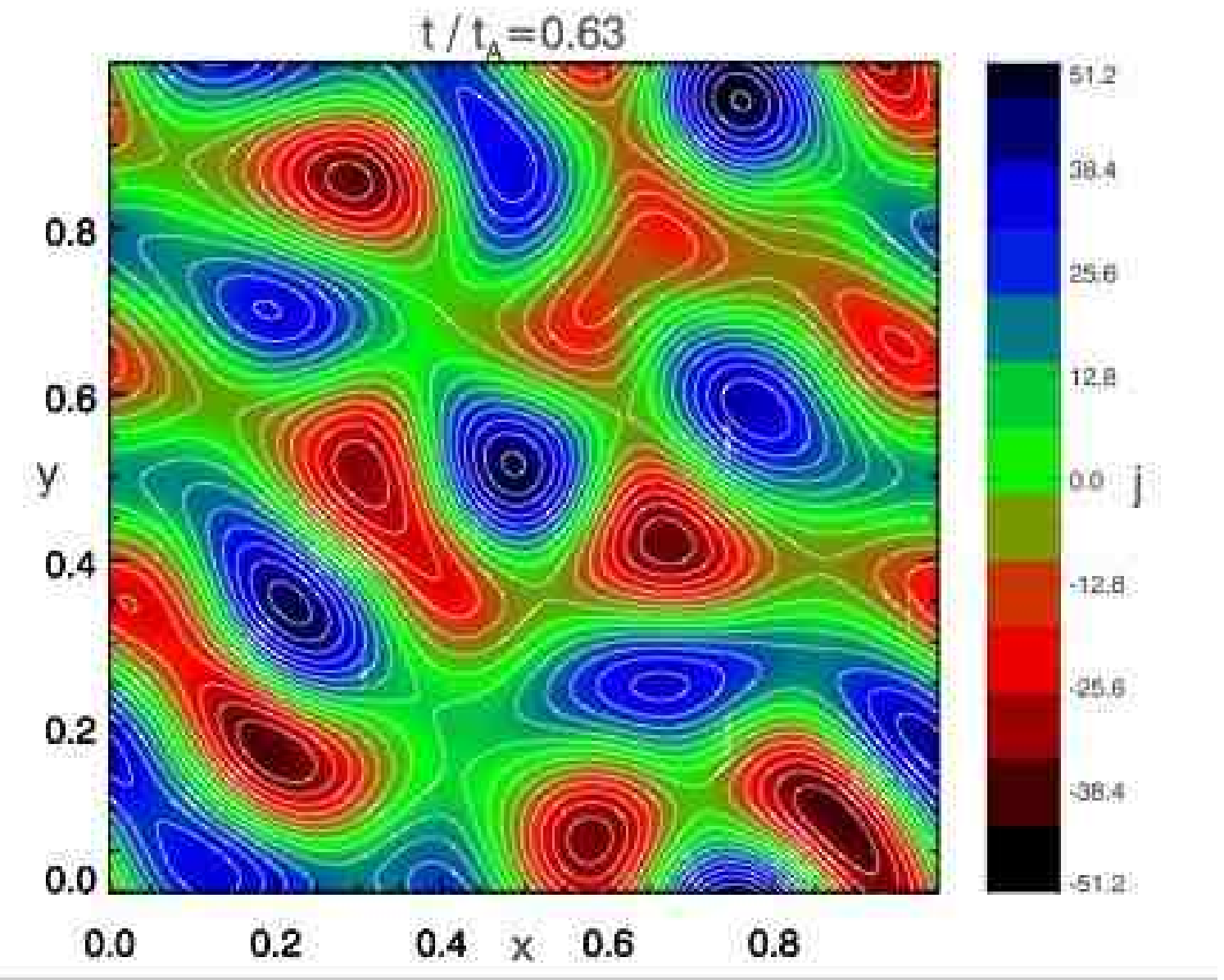}
               }\\[10pt]
      \subfloat[]{
               \label{fig:atsjfl:c}             
               \includegraphics[width=0.46\linewidth]{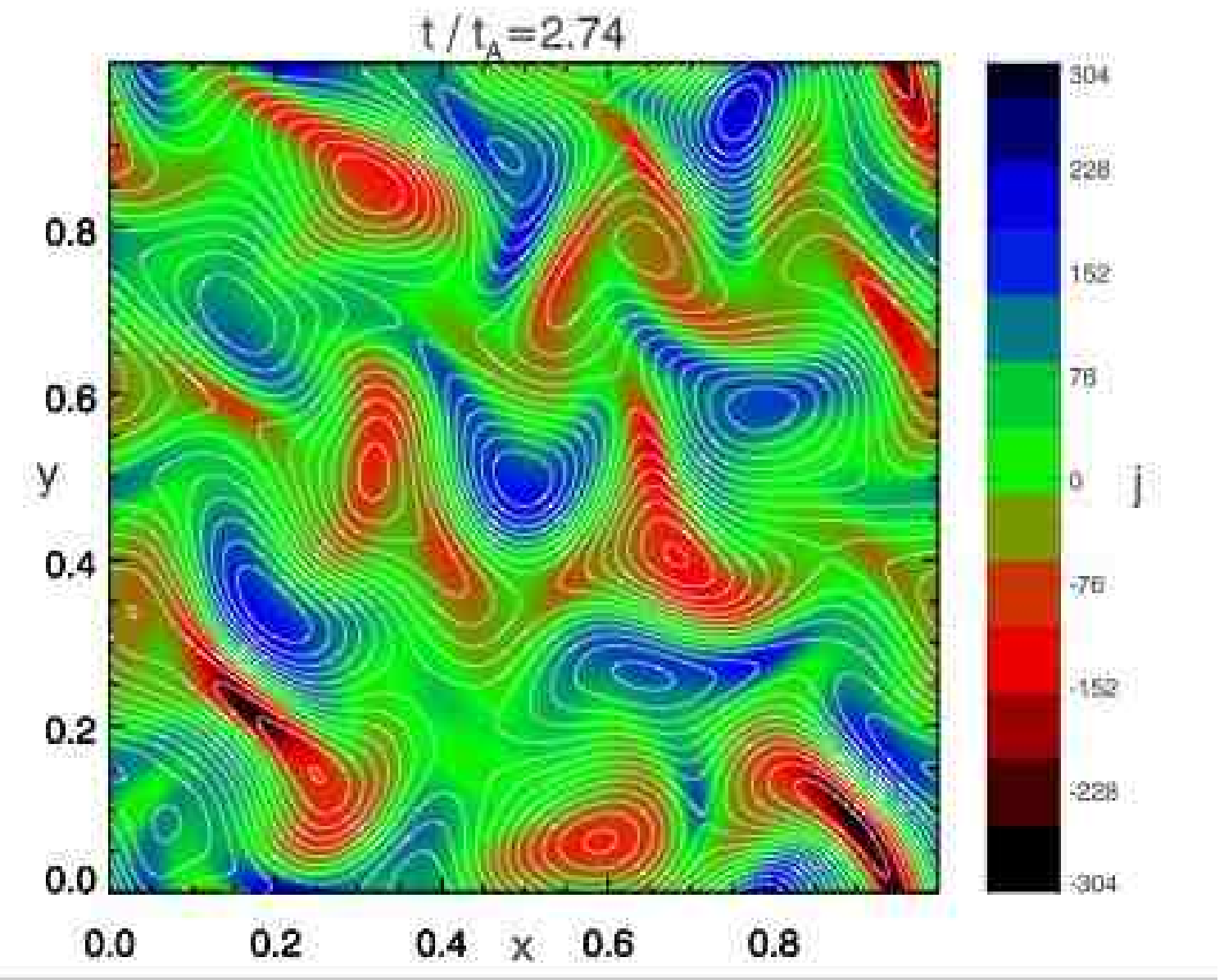}
               }
      \hspace{0.01\linewidth}
      \subfloat[]{
               \label{fig:atsjfl:d}             
               \includegraphics[width=0.46\linewidth]{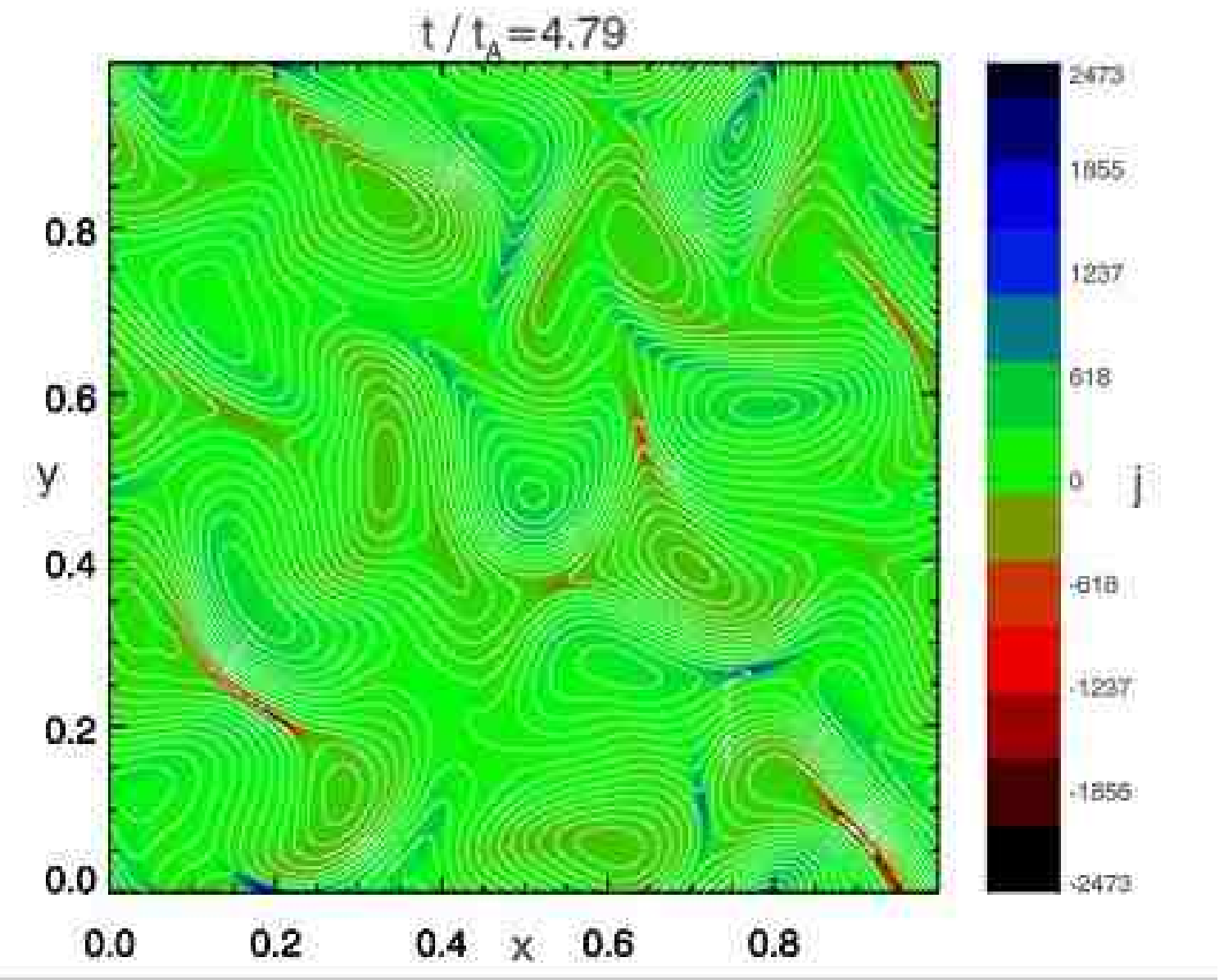}
               }\\[10pt]
      \subfloat[]{
               \label{fig:atsjfl:e}             
               \includegraphics[width=0.46\linewidth]{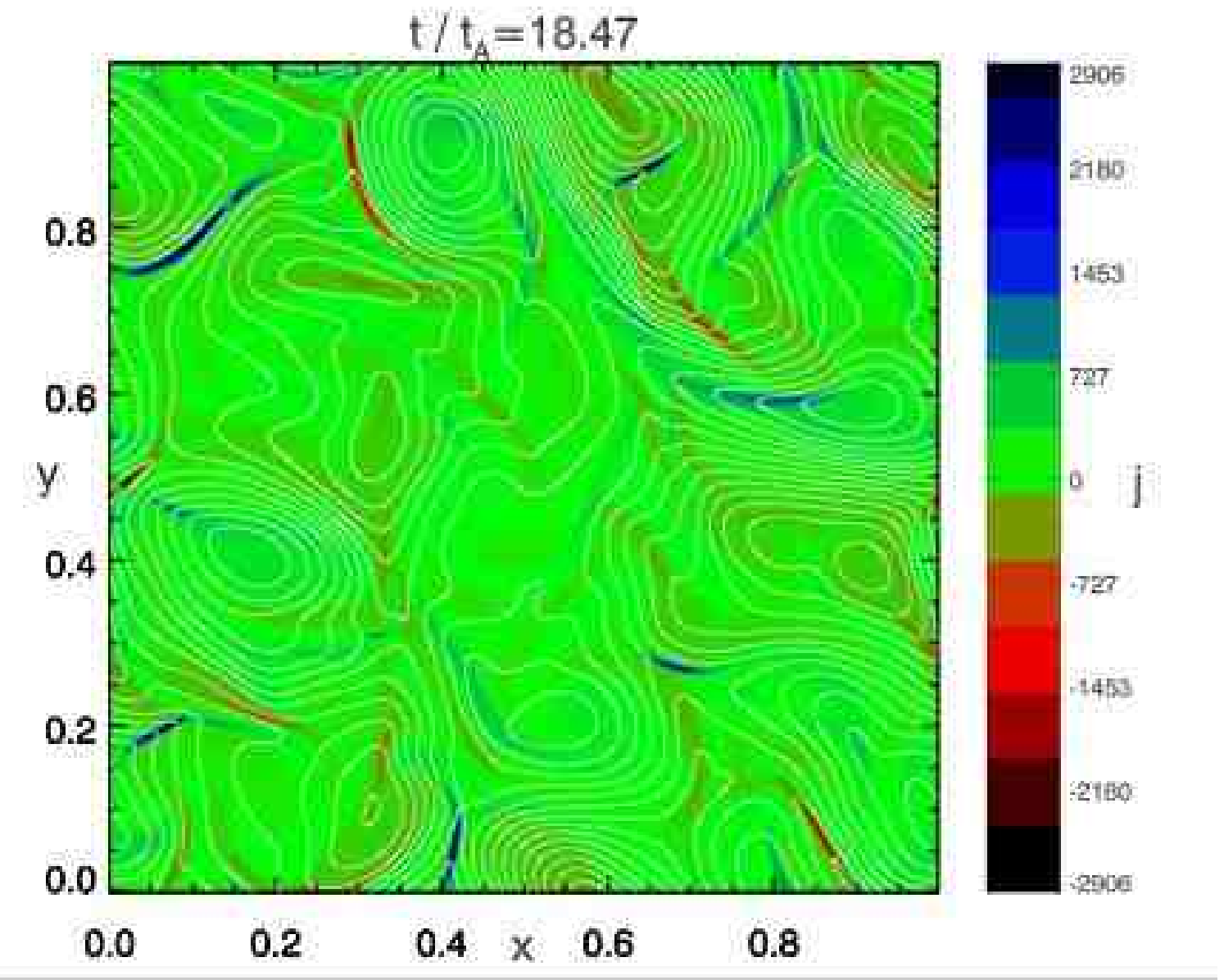}
               }
      \hspace{0.01\linewidth}
      \subfloat[]{
               \label{fig:atsjfl:f}             
               \includegraphics[width=0.46\linewidth]{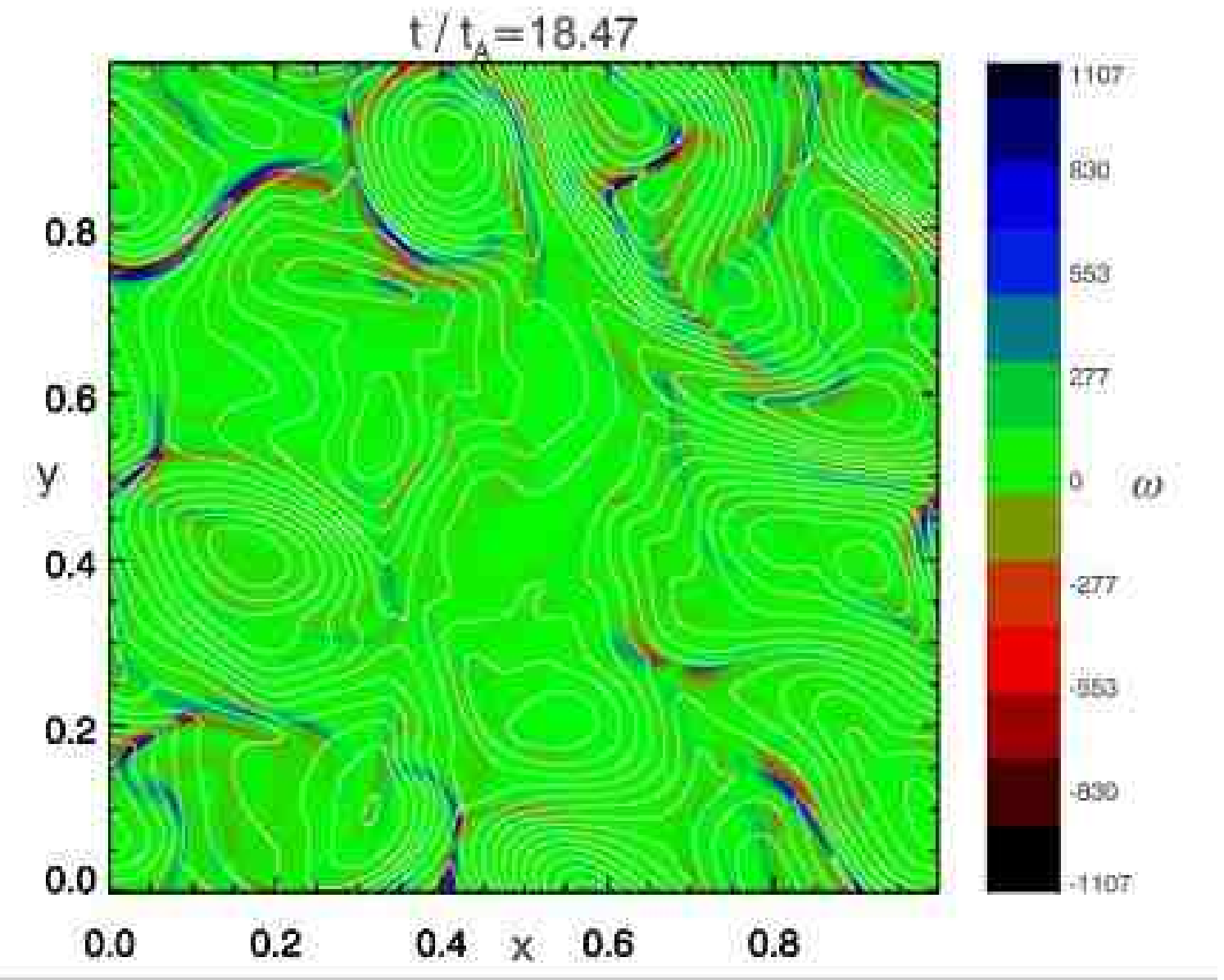}
               }
      \caption{\emph{Run A:} (a) Streamlines of the boundary velocity fields
               $\mathbf{u}_{\perp}^0 - \mathbf{u}_{\perp}^L$ constant
               in time. (b)-(e) Axial component of the current $j$ (in color) 
               and field-lines of the orthogonal magnetic field in the mid-plane 
               ($z=5$), at selected times covering the linear and nonlinear
               regimes up to $t = 18.47\, \tau_{\mathcal{A}}$. (f) Axial
               component of the vorticity $\omega$ (in color) and field-lines
               of the orthogonal magnetic field in the mid-plane at time
               $t = 18.47\, \tau_{\mathcal{A}}$.\\
               During the linear stage the orthogonal magnetic field is a mapping 
               of the boundary forcing (cfr.\ a and b). After the collaps
               of the large-scale currents (b, c, d), which in Fourier
               space correspond to a cascade of energy (see Figure~\ref{fig:emmod}),
               the topology of the magnetic field departs from the boundary velocity
               mapping and evolves dynamically in time (see movie).
               (e)-(f) Current sheets are embedded in quadrupolar vorticity 
               structure, a clear indication of nonlinear magnetic reconnection.
      \label{fig:atsjfl}}   
\end{figure}

\begin{figure}[p]
      \centering
      \subfloat[]{
               \label{fig:ats3D:a}             
               \includegraphics[width=0.47\linewidth]{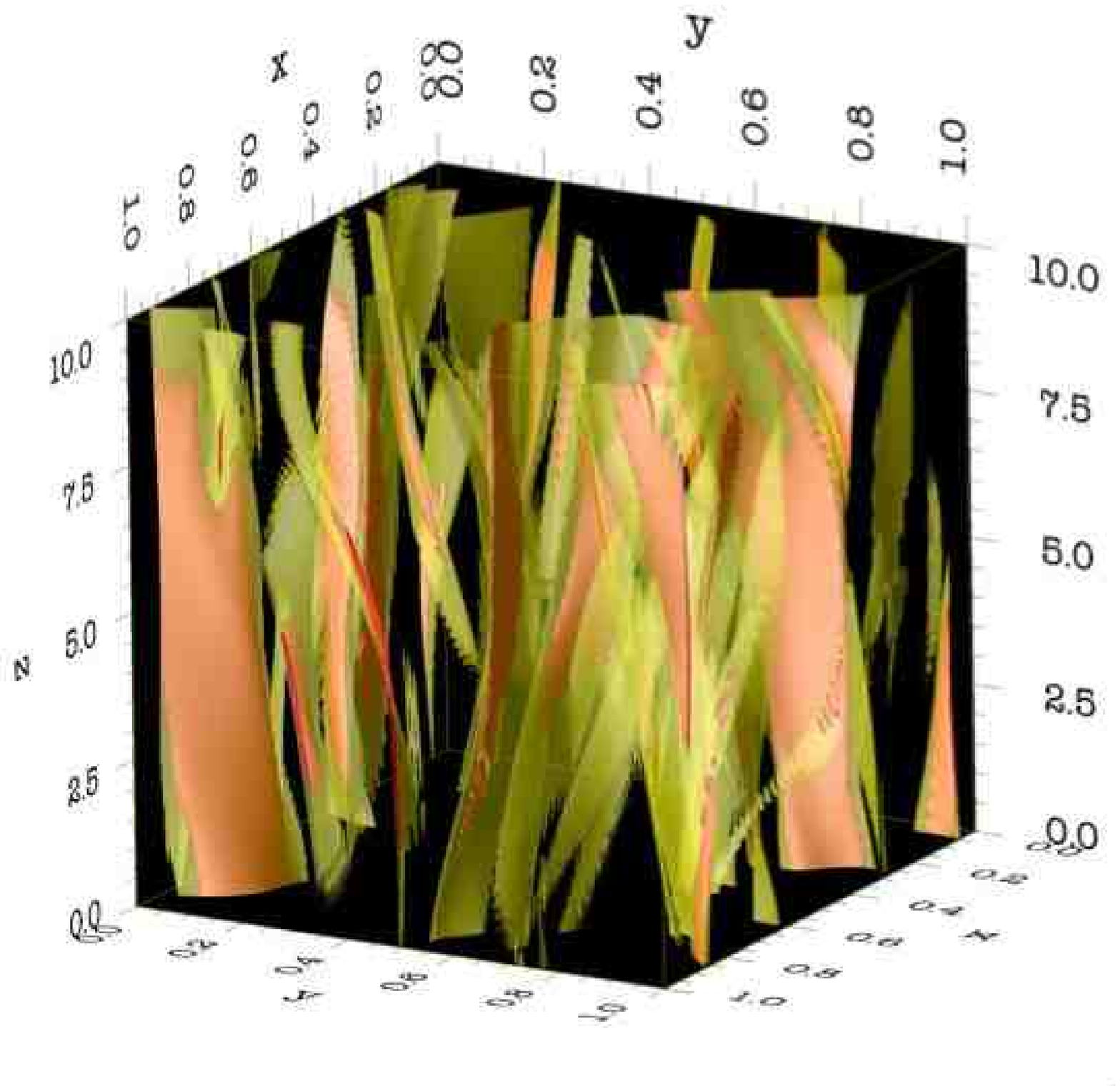}
               }
      \subfloat[]{
               \label{fig:ats3D:b}             
               \includegraphics[width=0.47\linewidth]{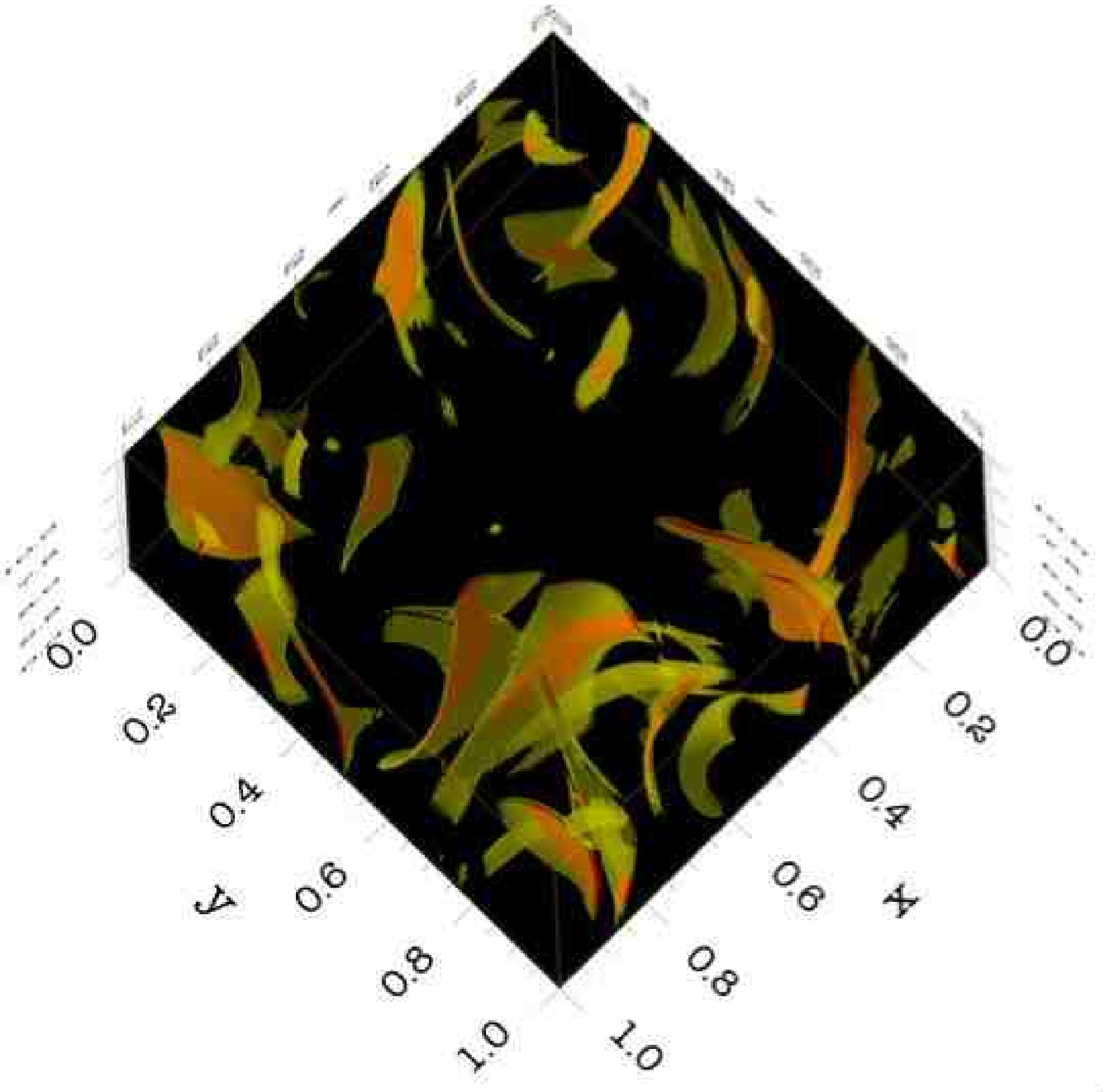}
               }\\[20pt]
      \subfloat[]{
               \label{fig:ats3D:c}             
               \includegraphics[width=0.47\linewidth]{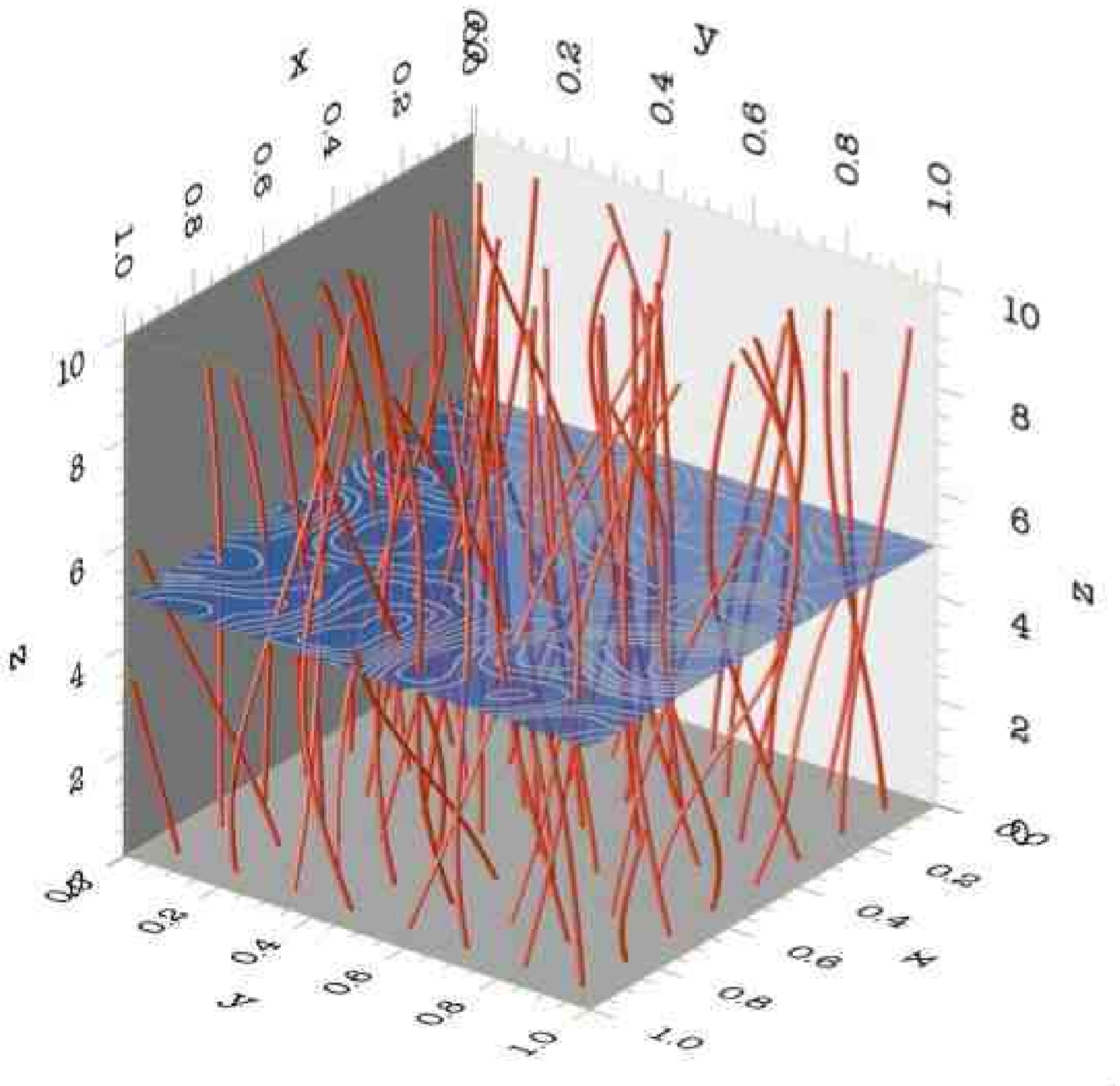}
               }
      \subfloat[]{
               \label{fig:ats3D:d}             
               \includegraphics[width=0.47\linewidth]{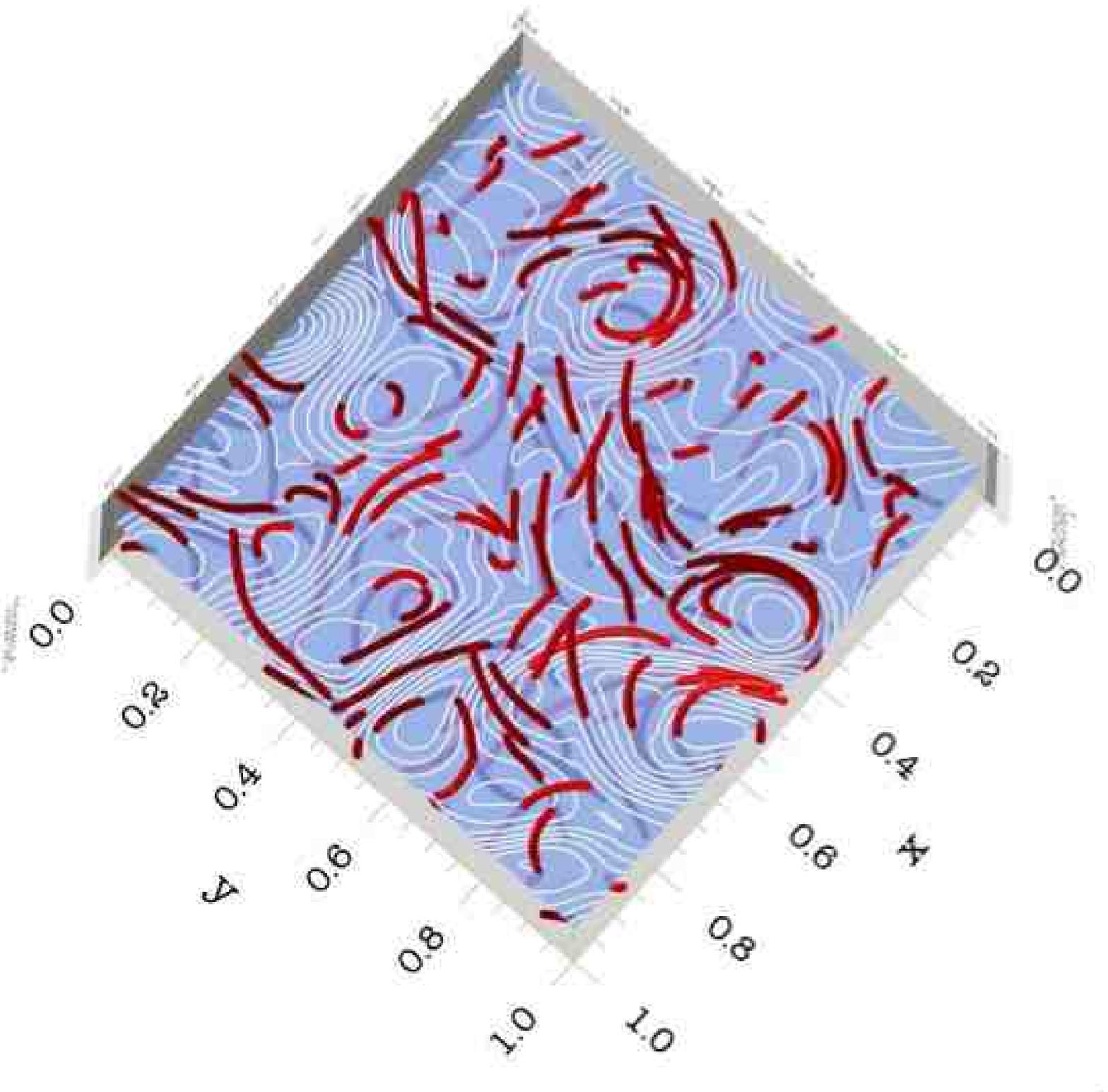}
               }
      \caption{\emph{Run A:} Side (a,c) and top (b,d) views of current
               sheets (a,b) and field lines of the total magnetic field
               (c,d) at time $t = 18.47\, \tau_{\mathcal{A}}$ (same time
               as in Figures~\ref{fig:atsjfl:e}-\ref{fig:atsjfl:f}). 
               For an improved visualization the box size has been rescaled,
               but the axial length of the computational box is 10 times
               longer that the perpendicular cross-section length.
               The rescaling of the box artificially enhances the structures
               inclination. To restore the original aspect ratio the box should
               be stretched 10 time along $z$.\\
               (a)-(b) Two isosurfaces of the squared current $j^2$. The isosurface
               at the value $j^2 = 2.8 \cdot 10^5$ is represented in partially 
               transparent yellow, while red displays the isosurface with 
               $j^2 = 8 \cdot 10^5$, well below the maximum value of the current
               at this time $j^2_{max} = 8.4 \cdot 10^6$. As typical of current
               sheets, isosurfaces corresponding to higher values of $j^2$ are
               nested inside those corresponding to lower values. For this 
               reason the red isosurface appears pink. Although from the side view
               the sheets appear space-filling, the top view shows that the filling
               factor is small.\\
               (c)-(d) Field-lines of the total magnetic field (orthogonal plus axial), 
               and in the mid-plane ($z=5$) field-lines of the orthogonal component of 
               the magnetic field.
      \label{fig:ats3D}}   
\end{figure}

\end{document}